\documentclass[useAMS]{mnras}
\usepackage{graphicx}
\usepackage{bm}
\usepackage{amssymb}
\usepackage{amsmath}
\usepackage{amsfonts}
\usepackage{mathrsfs}
\usepackage{epsfig}
\usepackage{mathtools}
\usepackage{enumerate}
\expandafter\ifx\csname package@font\endcsname\relax\else
 \expandafter\expandafter
 \expandafter\usepackage
 \expandafter\expandafter
 \expandafter{\csname package@font\endcsname}%
\fi

\newcommand{\ltsima} {$\; \buildrel < \over \sim \;$}
\newcommand{\gtsima} {$\; \buildrel > \over \sim \;$}
\newcommand{\lta} {\lower.5ex\hbox{\ltsima}}
\newcommand{\gta} {\lower.5ex\hbox{\gtsima}}

\newcommand{\lsim}{\raisebox{-.4ex}{$\stackrel{<}{\scriptstyle \sim}$}}
\newcommand{\gsim}{\raisebox{-.4ex}{$\stackrel{>}{\scriptstyle \sim}$}}

\newcommand{\RNum}[1]{\uppercase\expandafter{\romannumeral #1\relax}}

\begin{document}

\title[Spherical accretion in elliptical galaxies]
{Spherical accretion in giant elliptical galaxies: multi-transonicity, shocks, and implications on AGN feedback}

\author[S. Raychaudhuri et al.]
  {Sananda Raychaudhuri$^{1}$, Shubhrangshu Ghosh$^{1}$ \thanks{Email address: shubhrang.ghosh@gmail.com}\thanks{Present address: Department of Astronomy and Astrophysics, Tata Institute of Fundamental Research, Homi Bhabha Road, Colaba, Mumbai-400005, India}, Partha S. Joarder$^{1}$ \\ 
   $^{1}$ Centre for Astroparticle Physics and Space Science, Department of Physics, Bose Institute, \\ 
   Block EN, Sector V, Salt Lake, Kolkata, India 700091. \\ 
  }

\maketitle

\label{firstpage} 

\begin{abstract}
Isolated massive elliptical galaxies, or that are present at the center of cool-core clusters, are believed to be powered by hot gas accretion directly from their surrounding hot X-ray emitting gaseous medium. This leads to a giant Bondi-type spherical/quasi-spherical accretion flow onto their host SMBHs, with the accretion flow region extending well beyond the Bondi radius. In this work, we present a detailed study of Bondi-type spherical flow in the context of these massive ellipticals by incorporating the effect of entire gravitational potential of the host galaxy in the presence of cosmological constant $\Lambda$, considering a five-component galactic system (SMBH + stellar + dark matter + hot gas + $\Lambda$). The current work is an extension of Ghosh \& Banik (2015), who studied only the cosmological aspect of the problem. The galactic contribution to the potential renders the (adiabatic) spherical flow to become {\it multi-transonic} in nature, with the flow topology and flow structure significantly deviating from that of classical Bondi solution. More notably, corresponding to moderate to higher values of galactic mass-to-light ratios, we obtain Rankine-Hugoniot shocks in spherical wind flows. Galactic potential enhances the Bondi accretion rate. Our study reveals that there is a strict lower limit of ambient temperature below which no Bondi accretion can be triggered; which is as high as $\sim 9 \times 10^6 \, K$ for flows from hot ISM-phase, indicating that the hot phase tightly regulates the fueling of host nucleus. Our findings may have wider implications, particularly in the context of outflow/jet dynamics, and radio-AGN feedback, associated with these massive galaxies in the contemporary Universe. 
\end{abstract}

\begin{keywords}
accretion, accretion discs --- black hole physics ---  (cosmology:) dark energy --- (cosmology:) dark matter --- galaxies:active --- galaxies:jets
\end{keywords}	

\section{Introduction}

Evidences suggest that isolated massive elliptical galaxies or that are present at the central region of the cool-core clusters, are usually low-luminous, low-excitation radio galaxies (LERGs), that possess powerful radio-emitting jets emanating from their host active nuclei harbouring SMBHs (Hlavecek-Larrondo \& Fabian 2011; Janssen et al. 2012; see the recent review of Heckman \& Best 2014, hereinafter HB14). In these LERGs, most of the energy of their host active galactic nuclei (AGNs) is released in the bulk kinetic form through the radio-emitting jets (also referred to as jet-mode AGNs). These jets strongly interact with the ambient medium and inject considerable amount of energy into the surrounding gas, thereby heating the gas and prevent its radiative cooling (e.g., Churazov et al. 2002; McNamara \& Nulsen 2007; McCarthy 2010; Ishibashi et al. 2014). Indeed, X-ray observations show demonstrative evidence of the interaction between these jets and the interstellar medium (ISM) of their host galaxies, and/or with the intracluster medium (ICM) if the ellipticals reside at the centre of clusters (Fabian et al. 2003; McNamara et al. 2005; Allen at al. 2006; Nemmen at al. 2007, hereinafter N07; McNamara \& Nulsen 2007). This kinetic feedback (or radio-mode feedback) has been envisaged to account for the lack of star formation in these massive galaxies, and quenching of cooling flows in the isolated ellipticals or in the central region of the cool-core clusters (e.g., Fabian et al. 2003; Croton et al. 2006; Fabian 2012; Gaspari et al. 2012; Gaspari et al. 2013a; Croston et al. 2013), thereby preventing the collapse of the surrounding gas onto the centre of their host galaxies, inhibiting their growth (Narayan \& Fabian 2011, hereinafter NF11). Such a feedback is now thought to play a pivotal role in shaping the late evolution of these massive galaxies in the present-day Universe (e.g., Fabian 2012).  

A pertinent question now arises as to what fuels the central SMBHs in these LERGs. Unlike that of the luminous, high-excitation sources, whose central SMBHs are thought to be fuelled by the cold gas in a radiatively efficient accretion mode, the mode of fuelling for LERGs is less certain. Nonetheless, it is most likely that the hot gas may be the fuelling source for these jet-mode LERGs. The central SMBHs in the nuclei of these massive ellipticals (either in the context of isolated ellipticals or in the case of the ones present at the central region of the cool-core clusters) are immersed in a hot X-ray emitting gaseous medium, that provides an abundant gas source for their fuelling. Based on several evidences, it has been widely argued that host SMBHs in these massive galaxies are directly fuelled by these hot gases through a spherical/quasi-spherical accretion, in a radiatively inefficient mode (or hot mode accretion), at Bondi or at near-Bondi accretion rates (see di Matteo et al. 2003; Best et al. 2006; Hardcastle et al. 2007; Best \& Heckman 2012; HB14; Ineson et al. 2015). From a sample of nine nearby massive ellipticals, Allen et al. (2006), in fact, observed a tight correlation between their (computed) Bondi accretion rates ($\dot{M}_{B}$) and their total jet powers, suggesting that a Bondi-type spherical accretion can reasonably power these low-luminous AGNs. In reality, however, the in-falling gas would likely to retain some angular momentum, and under these circumstances, the nature of accretion would then be most appropriately described as a quasi-spherical, advection dominated accretion flow (ADAF) (Narayan \& Yi 1994, 1995; Bhattacharya et al. 2010; NF11; Yuan \& Narayan 2014; Ghosh 2017). Nonetheless, it has been pointed out by NF11 that to a great degree ADAF resembles a Bondi-type spherical accretion. In that paper the authors have, in fact, found that the jet power directly scales with Bondi accretion power rate, and argued that if the BH is fast rotating, a Bondi-type flow can support the observed jet power. Considering an ADAF, N07 has established an empirical relationship between the mass accretion rate (i.e., $\dot M_{\rm ADAF}$) and jet power for a sample of nine massive ellipticals [as used by Allen et al. (2006)], and found a same functional form as the correlation in Allen et al. (2006).  

An important aspect of (Bondi-type) hot mode accretion powering the LERGs, is that, the radio AGN feedback and the fuelling of host AGN gets tightly coupled. The radio-mode feedback keeps the 
surrounding gas hot. This hot phase then controls the fuelling of host AGNs, which in turn affects the radio AGN feedback itself. Thus a feedback loop is being maintained, enabling a `self-regulated' and an efficient feedback to occur in these galaxies. This feedback cycle helps in the maintenance of these massive galaxies in the contemporary Universe. The scenario then corresponds to a giant Bondi-type spherical accretion onto the central SMBH with the accretion-flow extending well beyond the Bondi radius, exceeding several hundred pc length-scale. Under these circumstances, it would be difficult to conceive a scenario in which the corresponding accretion flow towards the AGN is influenced by the gravitational field of the central SMBH alone; the galactic contribution to the gravitational potential needs to be considered in a more realistic modelling as was also pointed out earlier by other authors (e.g., NF11, Ishibashi et al. 2014). Moreover, for giant ellipticals present at the centre of galaxy clusters, like central dominant (CD) galaxies, whose radius may exceed several hundred kpc (e.g., Seigar et al. 2007), the host nucleus is fuelled directly from the hot phase of the ICM, and the length-scale of the corresponding accretion flow may then well extend beyond hundreds of kpc. In addition to the effect of the gravitational potential of galaxy, the influence of dark energy (e.g., Riess et al. 1998; Perlmutter et al. 1999) on the flow would then also be expected to become significant, at least at the outer regions of the flow. The simplest and most attractive candidate for the dark energy is the positive or repulsive cosmological constant ($\Lambda \sim$ 10$^{-52}$ m$^{-2}$), and the current paradigm of cosmology is based on $\Lambda$CDM model, where CDM refers to cold dark matter (e.g., Komatsu et al. 2011). Although the effect of $\Lambda$ is ignorable in the central regions of the galaxy, however, at length-scales $\gsim \, (50 - 100) \, {\rm kiloparsecs}$ from the nuclei of massive ellipticals, its effect may become non-negligible (e.g., Sarkar et al. 2014). 

To model the gravitational field of a massive elliptical galaxy, one needs to ascertain the total density or the mass distribution profile in that galaxy. Unlike in the case of spiral 
galaxies, in which, one can constrain the total mass distribution from the multicomponent modelling of their rotation curves, such an analysis is seemingly difficult in the context of ellipticals comprising mainly of four-mass-components (central SMBH, stellar, DM, and hot gas), as they have very little rotation (see for e.g., Mamon \& Lokas 2005b, hereinafter ML05b). ML05b have performed a detailed modelling of the four-component elliptical galaxy, describing the density profiles of stellar, DM, and hot gas components. In the context of massive ellipticals, if one incorporates the effect of $\Lambda$, the elliptical galaxy can then be described as a five-component gravitational system (SMBH + stellar + DM + hot gas + $\Lambda$). In recent times, Ghosh \& Banik (2015) (hereinafter GB15) have 
explored the possible impact of $\Lambda$ on the spherically symmetric accretion flows and found that $\Lambda$ suppresses Bondi accretion rate. However, in that work, the authors have ignored the explicit effect of the galactic potential in their calculations (also see Karkowski \& Malec 2013; Mach et al. 2013, in this context). On the other hand, Quataert \& Narayan (2000) had earlier endeavoured to investigate the Bondi-type spherical accretion in the gravitational field of a galaxy by considering a simplistic form of gravitational potential representing the rotation curve of the galaxy, however neglecting the effect of $\Lambda$, while Stuchl\'ik et al. (2016) have investigated the possible role of $\Lambda$ on spherically symmetric static mass configurations in the context of general-relativistic polytropes. 

In this paper, we endeavour to perform a detailed study of Bondi-type spherical accretion onto the central SMBH in the context of the above-mentioned five-component galactic system, to investigate how the galactic contribution to the gravitational potential in the presence of $\Lambda$ impact the dynamics of spherical accretion. Our work, described in this paper, is then a non-trivial extension of GB15, who studied only the cosmological aspect of the problem. Spherically symmetric flows have been extensively studied from multiple angles, not only in the context of accretion onto isolated BHs/compact objects (see the introduction of GB15), but also in the context of stellar wind theories (e.g., Axford \& Newman 1967; Summers 1980). Classical Bondi flow onto an isolated BH [which is inviscid and steady adiabatic (or polytropic)] is transonic in nature, always described by a single critical point. However there may be instances when this `criticality-condition' may be violated if the flow deviates from being strictly adiabatic, either with the appearance of subcritical points in the flow (e.g., Flammang 1982; Turolla \& Nobili 1989), or even obtaining more than one critical point in the spherical accretion (e.g., Chang \& Ostriker 1985; Turolla \& Nobili 1988; Nobili \& Turolla 1988; Nobili et al. 1991). In this work, we investigate in detail, the transonic behaviour of spherical accretion for a generic class of polytropic flows in the context of the five-component galactic system, to explore, whether the galactic contribution to the potential affects the `criticality-condition' of the classical Bondi solution. If such a criticality-condition is violated, this could substantially modify the properties of flow, and might significantly change the flow topology and structure relative to the classical Bondi solution, with the possibility of even occurrence of shocks in the flow. We examine such possibilities in the present study.

Among the various mass components in the elliptical galaxy, the precise nature of density or mass distribution profile for DM component is less definitive. Several DM density distribution models exist in the literature with the objective to describe $\Lambda$CDM halos in the dissipationless cosmological N body simulations (see ML05b; Merritt et al. 2006; Graham et al. 2006; Memola 2011; Stuchl\'ik \& Schee 2011, hereinafter SS11). Navarro, Frenk \& White (1995,1996) prescribed a double power-law DM density distribution profile with an outer slope of $ \simeq -3$ and an inner slope of $ -1$ (hereinafter NFW), obtained in large-scale high resolution dissipationless cosmological N-body simulations, and has been validated in many cosmological N-body simulations (see Lokas \& Mamon 2001, and references therein). Later simulation works have, however, revealed that the inner slope of the DM profile can actually be much steeper with its values lying within a range between $-3/2$ and $-1$ (e.g., Moore et al. 1999; Jing \& Suto 2000). A more general version of the NFW type DM density profile has been prescribed by Jing \& Suto (2000), with an outer slope of $ \simeq -3$ and an arbitrary inner slope of $- \gamma$, which is found to provide much better fit to simulated DM halos \footnote{For a detailed comparison of different DM models one can see Merritt et al. 2006.}. In the present study we adopt this generalized model of Jing \& Suto (2000) to describe the DM mass profile in our five-component galaxy. 

The rest of the paper is planned accordingly: In the next section, we formulate our five-component elliptical galaxy model. \S 3 briefly describes the hydrodynamical model for spherical accretion in the elliptical galaxy gravitational field, and the solution procedure. In \S 4 we analyse the transonic behaviour of spherical accretion in the elliptical galaxy gravitational field, and perform the global analysis of the parameter space. In \S 5, we study the fluid properties of spherical accretion in the context of our five-component elliptical galaxy model. In \S 6, we investigate how the galactic contribution to the potential influence the Bondi accretion rate. Finally, we end up in \S 7 with a summary and discussion. 

Before proceeding further, we furnish the values of the following cosmological quantities used in our study (e.g., ML05b; SS11): $\Lambda$ = 10$^{-52}$ m$^{-2}$; $\Omega_m$ (cosmological density parameter) = 0.3; $\Omega_b$ (baryon density parameter) = 0.041; $H_0$ (Hubble constant) = 100 $h$ km s$^{-1}$ Mpc$^{-1}$ = 70 $h_{70}$ km s$^{-1}$ Mpc$^{-1}$, $h$ = 0.7 (i.e. $h_{70}$ =1.0); ${\overline{\mathit{f}}_b}$ (mean baryon fraction of Universe) = $\Omega_b/\Omega_m \simeq 0.14$; ${\overline{\Upsilon}}_{B}$ (mass-to-light ratio of the Universe) = 390 $M_{\odot}/L_{\odot}$, where $M_{\odot}$ and $L_{\odot}$ are the solar mass and solar luminosity, respectively.

\section{Modelling elliptical galaxy gravitational field} 

Elliptical galaxies can be assumed to have nearly spherical mass distribution. In the presence of $\Lambda$, the spacetime geometry exterior to 
a static spherically symmetric mass distribution is Schwarzschild-de Sitter (SDS). The basic features of SDS spacetime have been extensively discussed in the 
literature (for e.g., see SS11; Sarkar et al. 2014; GB15); we do not repeat it here. Although the SdS geometry generally describes BH in a spatially inflated Universe, however, it can be applied to describe the spacetime outside any spherical or nearly spherical matter distribution, or even outside any mass distribution at length-scales where the deviation from spherical symmetry of the mass distribution can be ignored (e.g., SS11). Owing to the difficulty of studying complex astrophysical phenomena in the framework of full general relativity, many complex astrophysical phenomena has been studied through the use of pseudo-Newtonian potentials (PNPs), which are prescribed to approximately mimic corresponding GR effects, and has been extensively used in the astrophysical literature (e.g., Ghosh \& Mukhopadhyay 2007; Ghosh et al. 2014; Sarkar et al. 2014; Ghosh et al. 2015, 2016). Although few PNPs exist in literature that can well mimic SDS spacetime (Stuchl\'ik \& Kov\'a\v{r} 2008; Stuchl\'ik et al. 2009; Sarkar et al. 2014), here, we focus on the PNP prescribed in Stuchl\'ik et al. (2009), which is given by 

\begin{eqnarray}
\Psi_{\rm PN} \, (r) = - \frac{GM+ \frac{\Lambda c^2 \, r^3}{6}}{r-\frac{2GM}{c^2}- \frac{\Lambda r^3}{3}} \, , 
\label{1}
\end{eqnarray}
where $M$ is the gravitational mass of the distribution, $G$ and $c$ are the universal gravitational constant and speed of light, respectively. 
The subscript `PN' symbolizes `pseudo-Newtonian'. This PNP quite precisely reproduces salient features of corresponding SDS geometry, 
and has been used on number of occasions to study the effect of $\Lambda$ on relevant astrophysical phenomena (in local-scales) (Stuchl\'ik et al. 2009; SS11; GB15). Following the approach of SS11, we expand the PNP described in Eqn. (1), to obtain its Newtonian limit. The corresponding gravitational force in the Newtonian limit ignoring the higher order relativistic terms, is then given by 

\begin{eqnarray}
\mathscr{F}_{N} \, (r) =  \frac{GM}{r^2} - \frac{\Lambda c^2 \, r}{3} \, , 
\label{2}
\end{eqnarray}
where subscript `N' symbolizes `Newtonian'. In the right hand side of the above equation, the second term represents the form of gravitational force associated with $\Lambda$, whereas the first term describes the gravitational force associated with the spherical mass distribution; in the context of an elliptical galaxy, the mass distribution then comprises of four-mass components: BH, stellar, DM, and diffuse hot gas. In the framework of Eqn. (2), the net gravitational force associated with the elliptical galaxy, can then be simply expressed through a linear superposition of all the individual gravitational force terms associated with each different component of the elliptical galaxy, given by

\begin{align}
\mathscr{F}_{\rm Gal} \, (r) =  \mathscr{F}_{\rm BH} \, (r) + \mathscr{F}_{\rm star} \, (r) + \mathscr{F}_{\rm DM} \, (r) + \mathscr{F}_{\rm gas} \, (r) + \mathscr{F}_{\Lambda} \, (r) \, , 
\label{3}
\end{align}

where $\mathscr{F}_{\rm Gal}$ is the galactic force function. 
The corresponding galactic potential can then be written as $\Psi_{\rm Gal} \, (r) = \int \mathscr{F}_{\rm Gal} (r) \, dr$. 
The advantage of representing the gravitational effect of the galactic mass distribution in the presence of $\Lambda$ through the above additive fashion, is that, this would then enable one to study relevant astrophysical phenomena both inside the spherical galactic halo, as well as exterior to the galactic mass distribution. For more details, see SS11, although in a different context, where they used this approach to study the effect of $\Lambda$ on the motion of Small and Large Magellanic Clouds, in the gravitational field of the Milky Way. They found that the Newtonian limit of the PNP [described by Eqn. (1)] representing the effects of $\Lambda$ can be quite effectively used in the regions vicinity of the galactic disc and inside the galactic halo. Nonetheless, in the regions far outside the galactic halo the relativistic corrections are important and the full PNP then needs to be adopted. In the present work, as we are 
predominantly concerned with the fluid flow in the region either close to the galactic halo, or inside the galactic halo, we ignore the relativistic corrections 
(relating to $\Lambda$) and adopt the gravitational force function described by Eqn. (3), appropriate to our purpose. In this framework, the gravitational effect associated with the central SMBH is then represented through the Newtonian potential. This would then incorrectly describe the flow behavior in the vicinity of the BH, where the relativistic effects are important. However, as the galactic force function is expressed as the sum of individual gravitational force terms, we simply replace the Newtonian force term, and instead adopt the Paczy\'nski \& Wiita (1980) PNP, to capture the relativistic effect of the non-rotating BH; the corresponding force term given by 
\begin{eqnarray}
\mathscr{F}_{\rm BH} = \frac{G \, M_{BH}}{\left(r- {2GM_{\rm BH}}/{c^2} \right)^2} \, , 
\label{4}
\end{eqnarray}
where $r_g = {GM_{\rm BH}}/{c^2}$, $M_{\rm BH}$ is the BH mass. To determine the gravitational force functions associated with stellar mass distribution, DM, and hot gas mass, one needs to know their corresponding density or mass 
distributions. For details about the density distribution profiles associated with each of these components in the elliptical galaxy readers are advised to see ML05b, and references therein. Following ML05b, here, we obtain the gravitational force functions associated with these three components: stellar, DM and hot gas, which we furnish in the following subsections.

\subsection{Stellar mass gravitational field}

The stellar mass density distribution can be obtained by deprojecting the S\'ersic surface brightness profile (S\'ersic 1968); the S\'ersic law is found to well describe the surface brightness distribution of most of all the large elliptical galaxies. The S\'ersic profile follows the relation (ML05b)

\begin{eqnarray}
I(r) = I_{0} \, {\rm exp} \left[-\left(r/r_s \right)^{1/n}  \right] \, , 
\label{5}
\end{eqnarray}
where, $I$ is the surface brightness, $I_{0}$ is the normalization parameter, $r_s$ is the S\'ersic scale-radius and $n$ is the S\'ersic shape parameter or the `S\'ersic index'. The gravitational force associated with this distribution is then given by

\begin{eqnarray}
\mathscr{F}_{\rm star} \, (r) = \frac{G \, \mathit{f}_{\rm star} \, M_\nu}{r^2} \, \frac{\zeta \left[(3-\mu)n, \, \left(r/{r_s} \right)^{1/n} \right]}{\zeta \left[(3-\mu)n, \, \left({r_\nu}/{r_s} \right)^{1/n} \right]} \, , 
\label{6}
\end{eqnarray}
where $r_\nu$ represents the virial radius, which is defined to be the radius of that spherical region of the galaxy within which the mean total mass density is $200$ times the mean critical density of the Universe $\left(\rho_{\rm crit} \right)$, where $\rho_{\rm crit} = {3 H^2_0}/{8\pi G}$, this renders the virial radius $r_\nu \simeq 206.262 \, h^{-1}_{70} \, \left( {{h_{70} \, M_\nu} / {10^{12} \, M_\odot} } \right)^{1/3}$ Kpc. $M_\nu$ represents the total mass within the virial radius, $f_{\rm star}$ represents the stellar mass fraction within the virial radius. Here $\zeta (t,x)$ represents the standard incomplete gamma function, $\mu$ is related to S\'ersic shape parameter $n$ through the relation $\mu (n) \simeq 1.0 - 0.06097/n + 0.05463/n^2$. S\'ersic scale-radius $r_s$ is related to the effective 
(i.e., projected half-light) radius ${\mathscr R}_{\rm eff}$ through the relation $\mathscr{R}_{\rm eff} = b^{\mu} \, r_s$, with $b (n) \simeq 2n - 1/3 + 0.009876/n$. $\mathscr{R}_{\rm eff}$ and $n$ are given through the following relations based on the empirical fits to observations (see ML05b, and references therein). 

\begin{eqnarray}
\log \, \left(h_{70} \mathscr{R}_{\rm eff} \right) = 0.34 + 0.54 \, \log L_{10} + 0.25 \, \left(\log L_{10}\right)^2  \, ,
\label{7}
\end{eqnarray}
\begin{eqnarray}
\log \, n = 0.43 + 0.26 \, {\log L_{10}} - 0.044 \, \left(\log L_{10}\right)^2 \, , 
\label{8}
\end{eqnarray}
where $L_{10} = h^2_{70} \, L_B/{\left(10^{10} L_{\odot}\right)}$, $L_B$ is the luminosity of the galaxy in the blue-band. In Eqn. (7), 
$\mathscr{R}_{\rm eff}$ is measured in kpc.

\subsection{DM gravitational field}

A general version of the NFW type DM density distribution profile has been prescribed by Jing \& Suto (2000) which furnishes a double power-law with an outer slope of $\simeq -3$ and an arbitrary inner slope of $- \gamma$, with $1 \, \lsim \, \gamma \, \lsim \, 3/2$; the density profile is then given by 

\begin{eqnarray}
{\rho}_{_{\rm DM}}  \, (r) \propto \frac{1}{\left(r/r_d \right)^{\gamma} \, \left[1+ \left(r/r_d \right) \right]^{3- \gamma}} \, ,
\label{9}
\end{eqnarray}
where $r_d$ is some scale-radius which is related to the concentration parameter $c_o$ through the relation $c_o = {r_\nu}/r_d$. This above expression is a special case of a more generic $(\alpha, \beta, \gamma)$ model (see Merritt et al. 2006). With the choice of $(\alpha, \beta, \gamma) \equiv (1, 3, 1)$ one obtains the usual NFW model, and similarly with the choice of $(\alpha, \beta, \gamma) \equiv (1, 3, 3/2)$ one obtains the model described in Moore et al. (1999). Although the profile in Eqn. (9) has been referred to by various names in the literature, following ML05b, we refer to this profile as `JS' profile. In our study, we choose both the limits of $\gamma$; $\gamma = 1$, and $\gamma = 3/2$. The gravitational forces associated with this profile in the limits of $\gamma = 1$, and $\gamma = 3/2$, are then given by 

{\scriptsize
\begin{equation}
\mathscr{F}_{\rm DM} \, (r)= \begin{cases}
               \frac{G \, \mathit{f}_{\rm DM} \, M_{\nu}}{r^2} \, \frac{1}{\mathscr{G}(c_o)} \, \left[{\rm ln} \left(1+r/r_d \right) - \frac{r/r_d}{1+ r/r_d} \right] \, \, \, \, \, \, \, \, \, \,  \, \, \, \, \, \, \, \, \, \, \, \, \, \, \,  ({\rm NFW})   \, , \\ 
              \frac{G \, \mathit{f}_{\rm DM} \, M_{\nu}}{r^2} \, \frac{1}{\mathscr{G}(c_o)} \, \, 2 \left[{\rm sinh}^{-1} \, \left(\sqrt{r/r_d} \right) - \sqrt{\frac{r/r_d}{(1+r/r_d)}} \right] \, \, 
({\rm JS}- 3/2)  \, , 
             \end{cases}
\label{10}
\end{equation}
}
where $\mathit{f}_{\rm DM}$ represents the DM mass fraction within the virial radius, and

\begin{equation}
\resizebox{.497 \textwidth}{!}
{
$\mathscr{G}(c_o) = \begin{cases}
                {\rm ln} \left(1+c_o\right) - {c_o}/{\left(1+ c_o \right)}  \, \, \, \, \, \, \, \, \, \, \, \, \, \, \, \, \, \, \, \, \, \, \, \, \, \, ({\rm NFW})  \, , \\ 
                2 \left[{\rm sinh}^{-1} \, \left(\sqrt{c_o} \right) - \sqrt{{c_o}/{\left(1+c_o \right)}} \right] \, \,  ({\rm JS}- 3/2) \, , 
             \end{cases}$
}
\label{11}
\end{equation}
where the concentration parameter $c_o$ can be obtained from the $\Lambda$CDM simulations (see Jing \& Suto 2000) through the following fitting relations  

\begin{equation}
c_o \simeq \begin{cases}
                10.2 \, \left( {{h \, M_\nu} / {10^{12} \, M_\odot} } \right)^{-0.08} \, \, \, \, \, \, \, \, \, \, \, ({\rm NFW})  \, , \\ 
                4.9 \, \left( {{h \, M_\nu} / {10^{12} \, M_\odot} } \right)^{-0.13} \, \, \, \, \, \, \, \, \, \, \, \, \, ({\rm JS}- 3/2) \, . 
             \end{cases}
\label{12}
\end{equation}

\subsection{Hot gas mass gravitational field}

It is being generally assumed, that the hot X-ray emitting gas present in groups and clusters of galaxies, and in isolated early-type massive ellipticals, is in hydrostatic equilibrium in the overall gravitational potential (see Capelo et al. 2010; Capelo et al. 2012; and references in them). An isothermal $\beta$-model which provides a good fit to the X-ray observations of the early-type massive ellipticals, is being widely considered to describe the density distribution or the mass distribution of the hot gas in these systems (e.g., Brown \& Bregman 2001; ML05b, and references therein). In our analysis we adopt this $\beta$-model following ML05b, to describe the gas profile. The gravitational force associated with this model is given by

\begin{align}
\mathscr{F}_{\rm gas} \, (r) = \frac{G \, \mathit{f}_{\rm gas} \, M_\nu}{r^2} \, \frac{ \left[ \left(\frac{1}{3} \, \left({r}/{r_b} \right)^3 \, \right)^{-\delta} + \left(\frac{2}{3} \, \left({r}/{r_b} \right)^{\frac{3}{2}} \, \right)^{-\delta} \right]^{-1/{\delta}} }{ \left[ \left(\frac{1}{3} \, \left({r_\nu}/{r_b} \right)^3 \, \right)^{-\delta} + \left(\frac{2}{3} \, \left({r_\nu}/{r_b} \right)^{\frac{3}{2}} \, \right)^{-\delta} \right]^{-1/{\delta}} } 
\label{13}
\end{align}
where $\mathit{f}_{\rm gas}$ represents the mass fraction of the hot gas within the virial radius. $\delta = 2^{1/8}$, $r_b $ is some scale-radius which is related to the effective radius $\mathscr{R}_{\rm eff}$ through the relation $r_b \simeq {\mathscr{R}_{\rm eff}}/{q}$, with $q \simeq 10$ (ML05b). However, owing to the divergence of the associated gravitational potential at large radii, it has been assumed (ML05b) that beyond the virial radius (i.e, for $r > r_\nu$), the ratio of local baryon mass density to the local total mass density of the galaxy is approximately equal to the mean baryon fraction of the Universe. This then leads to the following definition of gravitational force associated with the hot gas for $r > r_\nu$,    

\begin{align}
\mathscr{F}_{\rm gas} \, (r) \, \, \vert_{r \, > \, r_\nu} = \frac{G \, M_{\nu}}{r^2} \, \left(\mathit{f}_{\rm gas} -  \mathit{f}_{\rm DM} \, \frac{\overline{\mathit{f}}_b}{1 - \overline{\mathit{f}}_b}   \right) \, + \,  \frac{\overline{\mathit{f}}_b}{1 - \overline{\mathit{f}}_b} \,  \mathscr{F}_{\rm DM} \, (r)  \, 
\label{14}
\end{align}
where the stellar mass contribution has been neglected, as the stellar luminosity almost converges at the virial radius. 
\vspace{3mm}

\par\noindent\rule{85mm}{0.7pt}
\vspace{0.0mm}

For the quantitative estimate of the force function $\mathscr{F}_{\rm Gal} \, (r)$ of the elliptical galaxy [given by Eqn. (3)], one needs to provide the values for the 
following quantities: $L_B$, $M_\nu$, $M_{\rm BH}$, $\mathit{f}_{\rm star}$, $\mathit{f}_{\rm DM}$, and $\mathit{f}_{\rm gas}$. We define mass-to-light ratio bias $\mathit{b}_{\Upsilon} ={\Upsilon_B}/{\overline{\Upsilon}}_{B} $, i.e., the ratio of the galactic mass-to-light ratio in the blue-band within the virial radius 
($\Upsilon_B$) to the universal mass-to-light ratio (${\overline{\Upsilon}}_{B}$). Note that we always express mass-to-light ratios in the units of $M_{\odot}/L_{\odot}$. 
Following ML05b, (also see, Capelo et al. 2010), $M_\nu$ can be written as $M_\nu = \mathit{b}_{\Upsilon} \, {\overline{\Upsilon}}_{B} \, L_B$. In expressing the above relation for $M_\nu$ it is being considered that the luminosity of the galaxy almost converges at the virial radius. 

A common widely used technique (although through indirect means) to estimate the central SMBH mass in massive early type galaxies (see HB14), is based on a fit to the relationship between $M_{\rm BH}$ and the stellar velocity dispersion ($\sigma$) of the surrounding galactic bulge (e.g., Ferrarese \& Merritt 2000; Gebhardt et al. 2000; Tremaine et al. 2002; Graham \& Scott 2013; McConnell \& Ma 2013), or using the correlation between $M_{\rm BH}$ and the stellar mass of the bulge ($M_{\rm bulge}$) (e.g., Marconi \& Hunt 2003, H\"aring  \& Rix 2004). The estimated BH mass from the fitting relation of  \& Rix (2004) is quite close to that obtained from the relation given in McConnell \& Ma (2013). In the present study, we choose the fitting relation of Haring \& Rix (2004) to estimate the central SMBH mass, given by
\begin{align}
\log \, \left(M_{\rm BH}/M_{\odot} \right) =  \left(8.20  \pm 0.10 \right) \, + \, \left(1.12  \pm 0.06 \right) \, \nonumber \\
\times \, \log \, \left(M_{\rm bulge}/10^{11} M_{\odot} \right),
\label{15}
\end{align}
where $M_{\rm bulge}$ can be written as $M_{\rm bulge} = \mathit{f}_{\rm star} \, M_\nu$. Following e.g., ML05b, the virial stellar mass fraction $\mathit{f}_{\rm star}$ can be written as 

\begin{eqnarray}
\mathit{f}_{\rm star} = \frac{\Upsilon_{\star\, B}}{{\mathit{b}}_{\Upsilon} {\overline{\Upsilon}}_{B}} \, , 
\label{16}
\end{eqnarray} 

where $\Upsilon_{\star\, B}$ is the stellar mass-to-light ratio in the blue-band. We define the baryon fraction bias $\mathit{b}_b = {\mathit{f}_b}/{\overline{\mathit{f}}_b}$, i.e, the ratio of baryon mass fraction within the virial radius ($\mathit{f}_b$) to the mean baryon fraction of the Universe ($\overline{\mathit{f}}_b$). One can then write $\mathit{f}_{\rm DM} = 1- \mathit{b}_b \, \overline{\mathit{f}}_b$, and $\mathit{f}_{\rm gas} = 1 - \left(\mathit{f}_{\rm star} + \mathit{f}_{\rm DM} \right) $, where it has been assumed that the fraction of BH mass to the stellar mass 
is much less than unity (e.g., ML05b, Capelo et al. 2010). 

In our present study, following for e.g., ML05b, Mamon \& Lokas (2005a), Capelo et al. 2010, we adopt the fiducial value for $L_B \simeq 2 \times 10^{10} \, L_{\odot}$, which has been used 
as a standard value for this parameter by these authors. It has been argued that $\Upsilon_{\star\, B}$ of elliptical galaxies lay roughly in the range from $\sim$ 5 - 8 (e.g., Mamon \& Lokas 2005a). Following for e.g., ML05b, Capelo et al. (2010), in the present work we adopt the mean value of $\Upsilon_{\star\, B} = 6.5$. Had the Universe been unbiased, $\Upsilon_B$ would have been always equals to 
${\overline{\Upsilon}_B} (= 390)$. For a comprehensive analysis of the dynamical behaviour of the flow relevant for our purpose, following ML05b, we adopt four fiducial values for the galactic mass-to-light ratio ($\Upsilon_B$) within a wide range $\sim (14-390)$, as depicted in Table 1. The lower limit of this range corresponds to the typical choice made by ML05b for the special case of negligible or no DM scenario. In fact, for NGC 821, an intermediate luminous elliptical, which has been predicted to have little or no DM content, Romanowsky et. al (2003) have estimated the value of $\Upsilon_B$ in the range $\sim 13 - 17$. Predictions from many previous studies (for e.g., Benson et al. 2000; Marinoni \& Hudson 2002; Yang et al. 2003) indicate that for galaxy-sized virialized halos, the value of 
$\Upsilon_B$ could exceed $300$, however, unlikely to surpass ${\overline{\Upsilon}}_B$. Following ML05b, and without loss of generality, we adopt $\Upsilon_B \simeq {\overline{\Upsilon}}_B = 390$ as our typical upper bound limit. Following Marinoni \& Hudson (2002) and Yang et al. (2003), the particular value of $\Upsilon_B = 100$ was adopted as a standard value in ML05b (also see Capelo et al. 2010), which we also adopt in our present study. The value of $\Upsilon_B  =  33$ was considered earlier by ML05b based on the findings of Romanowsky et al. (2003) for the nearby giant elliptical galaxy, NGC 3379; this particular value of $\Upsilon_B  =  33$ corresponds to the scenario of low content of DM. In this paper we examine our numerical results for all the above values of $\Upsilon_B$. For $\Upsilon_B = 100$, one can assume that the baryon fraction would remain unaffected, with $\mathit{f}_b = \overline{\mathit{f}}_b \simeq 0.14$ (ML05b). For the case with $\Upsilon_B  =  33$, we adopt similar value for the gas mass-to-stellar mass ratio as that obtained for $\Upsilon_B  =  100$ (ML05b). For the above values of $\Upsilon_{B}$, using Eqn. 16, and the relations furnished in the previous paragraph, one may easily estimate $\mathit{f}_{\rm star}$, $\mathit{f}_{\rm DM}$, and $\mathit{f}_{\rm gas}$. In Table 1, we enlist in detail, the relevant values for different physical quantities used to compute the galactic force function $\mathscr{F}_{\rm Gal} \, (r)$. For clarity, in Table 2, we enlist the various parameters used in this section. In the next section, we briefly describe the hydrodynamical model for spherical accretion flow in the elliptical galaxy gravitational field.

\begin{table}
\centering
\centerline{\bf Table 1}
\vspace{0.3cm} 
\centerline{\large $L_B = 2 \times 10^{10} \, L_{\odot} $,  $\Upsilon_{\star\, B} = 6.5 \, M_{\odot}/L_{\odot}$ } 
\begin{tabular}{|c|c|c|c|c|c|c|c|c|c}

\hline
$\Upsilon_B \, \left({M_\odot}/{L_\odot} \right)$ & ${\mathit{b}}_{\Upsilon}$ & $\mathit{f}_{\rm star}$ & $\mathit{b}_b$ & $\mathit{f}_{\rm DM}$ & $\mathit{f}_{\rm gas}$  \\  [1ex]
\hline
390  & 1.0  & 0.0167 & 1.0  & 0.86 & 0.1233   \\  [1ex]
\hline
100 & 0.25641 & 0.065 & 1.0 & 0.86 & 0.075 \\  [1ex]
\hline
33  &  0.08462 &0.1969 & 3.0307 & 0.5757 & 0.2272    \\  [1ex]
\hline
14  & 0.0359 & 0.4643 & 7.1429  & 0.0  & 0.5357     \\  [1ex]
\hline
\end{tabular}

{ \footnotesize For the choice of these fiducial parameters see Mamon \& Lokas (2005b) and references therein.}
\end{table}

\begin{table*}
\centering
{{\bf Table 2:}} {Glossary of the various parameters used in section 2.}
\begin{tabular}{|c|c|c|c|c|c|c|c|c|c}
\hline
$H_0$ & Hubble constant  & $ $ & $ $ & $L_B$ &  Luminosity of the galaxy in the blue-band \\  [0.1ex]
$h$ & Reduced Hubble constant & $ $ & $ $ & $\Upsilon_{\star\, B}$ &   Stellar mass-to-light ratio in the blue-band \\ [0.1ex]
$\Lambda$ & Cosmological constant & $ $ & $ $ & ${\overline{\Upsilon}}_{B}$ & Mass-to-light ratio of the Universe\\  [0.1ex]
$\rho_{\rm crit}$ & Mean critical density of the Universe & $ $ & $ $ & ${\Upsilon}_{B}$ & Galactic mass-to-light ratio \\  [0.1ex] 
$r_\nu$ & Virial radius & $ $ & $ $ & $\mathit{b}_{\Upsilon}$ & Ratio of ${\Upsilon}_{B}$ to ${\overline{\Upsilon}}_{B}$\\  [0.1ex]
$M_\nu$ & Virial mass  & $ $ & $ $ & ${\overline{\mathit{f}}_b}$ & Mean baryon fraction of Universe\\  [0.1ex]
$\mathscr{R}_{\rm eff}$ & Projected half-life radius & $ $ & $ $ &  $\mathit{b}_b$ & Baryon fraction bias\\  [0.1ex]
$r_s$ & S\'ersic scale-radius & $ $ & $ $ & $f_{b}$ & Baryon mass fraction within the virial radius\\  [0.1ex]
$n$   & S\'ersic shape parameter & $ $ & $ $ & $f_{\rm star}$ & Stellar mass fraction within the virial radius\\  [0.1ex]
$c_o$ & Concentration parameter & $ $ & $ $ &$f_{\rm DM}$ & Dark matter mass fraction within the virial radius\\  [0.1ex] 
$\gamma$ & - Inner slope of DM profile & $ $ & $ $ &$f_{\rm gas}$ & Hot gas mass fraction within the virial radius\\ [0.1ex]
\hline
\end{tabular}
\end{table*} 

\section{Hydrodynamical model for spherical accretion in the elliptical galaxy gravitational field and the solution procedure} 

Throughout our analysis, we actually express the radial coordinate $r$ in units of $r_g = {GM_{\rm BH}}/c^2$. The radial velocity of the flow $v_r$ is 
expressed in units of speed of light $(c)$. As the accretion flow is considered to be spherically symmetric onto a nonrotating 
BH traversing through the gravitational field associated with the different components of the elliptical galaxy, the dynamical flow variables are expressed only in functions of $r$, and as usual, flow is assumed to be inviscid in nature. Hence heat generation and radiative heat loss from the system are neglected. We consider a generic polytropic flow consisting of fully ionized Hydrogen and Helium plasma having galactic abundance. The equation of state of a polytropic flow follows a relation $P = K \rho^{1+1/N}$, where $P$ is the gas pressure, $\rho$ is the density, 
of the accreting plasma, $N$ is the polytropic index, and $K$ carries the information of the entropy of the flow. The sound speed $c_s$ follows the relation $c_s^2 = {\left(1+1/N \right) P}/\rho ={\left(1+1/N \right) k_B T}/{\mu m_p}$, where $k_B$ is the Boltzmann constant, $T$ the temperature of the flow, $m_p$, the usual mass of proton, and $\mu = 0.592$ is the mean molecular weight for the galactic abundance of Hydrogen and Helium with Hydrogen mass fraction $X = 0.75$. We also assume a quasi stationary accretion flow. A point needs to be noted here: In the usual viscous accretion flow theory, viscosity is generally introduced in the system through the `$r\phi$' component of viscous stress tensor in the angular momentum balance equation, which is required for the outward transport of angular momentum. The other components like `$rr$' component of viscous stress tensor which ought to have appeared in the radial momentum equation is neglected given the fact that $v_r$ is much less than $c_s$. It has been, however, pointed out by NF11, that for spherical type accretion flows in which case $v_r$ is sufficiently large, the `$rr$' component of viscous stress tensor could not altogether be wished away in the radial momentum equation, and hence ought to be incorporated in self consistent modelling of spherical accretion flows. Nonetheless, in the present work, we neglect this term in the radial momentum equation following the similar argument given in NF11. It needs to be noted, however, that the effect of viscosity on the spherically symmetric transonic flows have been discussed in the literature (e.g., Ray 2003; Turolla \& Nobili 1989; Summers 1980, and references therein). It would be then quite interesting to analyse the effect of this viscous stress term on the spherical accretion solution in the context of our system, which is left for future work. \\

The basic conservation equations for an inviscid steady state spherical accretion flow in the presence of elliptical galaxy gravitational field are then given by \\

(a) Mass conservation equation: 
\begin{eqnarray}
\dot M =  - 4 \pi r^2 \rho \, v_r \, ,  
\label{17}
\end{eqnarray}
where, $\vert \dot M \vert$ is the mass accretion rate of the flow. \\
(b) Radial momentum conservation equation: \\
\begin{eqnarray}
v_r \frac{d v_r}{dr} + \frac{1}{\rho} \frac{dP}{dr} 
+ \mathscr{F}_{\rm Gal} (r) = 0 \, . 
\label{18}
\end{eqnarray}
Following the procedure adopted in previous works for studying transonic accretion flow around BH/compact 
objects (e.g., Chakrabarti 1990, 1996; Narayan et al. 1997; Mukhopadhyay \& Ghosh 2003; GB15) we combine Eqns. (17) and (18) and obtain 
\begin{eqnarray}
\frac{d v^2_r}{dr} = 2 v^2_r \frac{\left[\frac{2 c^2_s}{r} - \mathscr{F}_{\rm Gal} (r) \right]}{v^2_r -c^2_s} = \frac{N1 \, (v_r, c_s, r)}{D \, (v_r, c_s)}  \, 
\label{19}
\end{eqnarray}
and 
\begin{eqnarray}
\frac{d c^2_s}{dr} = - \frac{1}{N} c^2_s \frac{\left[\frac{2 v^2_r}{r} 
- \mathscr{F}_{\rm Gal} (r) \right]}{v^2_r -c^2_s}  = \frac{N2 \, (v_r, c_s, r)}{D \,(v_r, c_s)} \, , 
\label{20}
\end{eqnarray}
respectively. Here `$N1 \, (v_r, c_s, r)$' and `$N2 \,(v_r, c_s, r)$' represent the numerators of Eqns. (19) and (20) respectively, and `$D \, (v_r, c_s)$' represents the 
denominator which is identical in both the Eqns. (19) and (20). The dynamics of the fluid 
flow is then governed by the Eqns. (19) and (20). The corresponding 
equations show that to guarantee a smooth solution, the following identity 
$N1=N2=D=0$ should be satisfied at a particular radius called `critical radius', or in this case the `sonic radius' ($r_c$). Thus at critical radius or the sonic radius, i.e., at $r_c$, the above criteria renders 
\begin{eqnarray}
v_{\rm rc} = c_{\rm sc} =  \sqrt{\frac{ r_c \, \mathscr{F}_{\rm Gal} (r_c)}{2}}\, .
\label{21}
\end{eqnarray}
where, $v_{\rm rc}$ and $c_{\rm sc}$ are the radial velocity and the sound speed at the sonic location ($r_c$). 
The existence of the sonic location(s) determines the transonic nature of the accretion flow. In order to 
describe the transonic behaviour of the accreting system, and to analyze the 
flow velocity profiles, one needs to find the solutions of the Eqns. (19) and (20). 
At sonic location(s), $\left.\frac{d v^2_r}{dr} \right\vert_{r_c} = \left.\frac{d c^2_s}{dr} \right\vert_{r_c} = \frac{0}{0}$. Hence we apply 
l'Hospital's rule to Eqns. (19) and (20) respectively. Both these equations at sonic location are then given by  
\begin{eqnarray}
\left.\frac{d v^2_r}{dr} \right\vert_{r_c} = \frac{- \, \mathcal{Q}_1 \, \pm \, \sqrt{{\mathcal{Q}_1}^2 - 4 \, \mathcal{P}_1 \mathcal{R}_1 }}{2 \mathcal{P}_1 } 
\label{22}
\end{eqnarray}
and 
\begin{eqnarray}
\left.\frac{d c^2_s}{dr} \right\vert_{r_c} = - \frac{1}{N} \left[\mathscr{F}_{\rm Gal} (r_c) 
+ \frac{1}{2} \left.\frac{d v^2_r}{dr} \right\vert_{r_c} \right] \, , 
\label{23}
\end{eqnarray}
respectively, where 

\begin{equation}
\begin{rcases*}
{\mathcal{P}}_1 =   \frac{1}{2} \, (2N + 1)/N , \\ 
{\mathcal{Q}}_1 = \frac{1}{N} \left[\frac{2 v^2_{rc}}{r_c} + \mathscr{F}_{\rm Gal} \, (r_c) \right] , \\
{\mathcal{R}}_1  =  2 v^2_{rc} \left[\frac{2 c^2_{sc}}{r^2_c} + \frac{1}{N}\frac{2}{r_c} \mathscr{F}_{\rm Gal} (r_c)   + \left.\frac{d \mathscr{F}_{\rm Gal} (r)}{dr}\right\vert_{r_c}       \right] .          
\end{rcases*}
\label{24}
\end{equation}

Dynamically, Bondi solution is transonic in nature comprising of two solutions: accretion and wind, described by Eqns. (19) and (20). 
`$-$' sign in the `$\pm$' in Eqn. (22) represents the accretion solution, whereas the `$+$' sign represents the wind solution.  
At sonic location $r_c$, both the flow velocity $v_r$ and sound speed $c_s$ correspond to both accretion and 
wind solutions, coincide. Integrating Eqns. (18) and (17) we can write the specific energy and entropy of the flow at 
sonic location (see GB15, and references therein), given by 
\begin{eqnarray}
E_c = \left(\frac{1}{2} + N \right) \, \frac{ r_c \, \mathscr{F}_{\rm Gal} (r_c)} 
{2}  + \Psi_{\rm Gal} \, (r_c) 
\label{25}
\end{eqnarray}
and 
\begin{eqnarray}
\dot {\mathcal{M}}_c =  4 \pi r^2_c\left[ \frac{ r_c \, \mathscr{F}_{\rm Gal} (r_c)} 
{2} \right ]^{(2N+1)/2} \, ,
\label{26}
\end{eqnarray}
respectively, where the specific entropy of accretion flow is simply related to the mass accretion rate through the relation $\dot {\mathcal{M}} = 
(\Gamma K)^N {\dot M}$ (see, Chakrabarti 1989b, 1990). Here, $\Gamma$ is the adiabatic index which 
is related to the polytropic index through the relation $N = 1/(\Gamma -1)$. In case of locally adiabatic flow $\Gamma$ varies between $4/3$ for a relativistic flow to $5/3$ for a nonrelativistic flow. 
While $\dot M$ is always a conserved quantity for the particular accretion flow in the steady state, however, in general, 
$\dot {\mathcal{M}}$ may not always be the so, owing to the fact that $\dot {\mathcal{M}}$ contains $K$ which carries the information of the entropy of the flow. For a dissipative system $\dot {\mathcal{M}}$ is not a conserved quantity, however for a nondissipative system $\dot {\mathcal{M}}$ remains conserved in the flow, unless a shock appears in the system. To solve the 
dynamical equations and study the global family of solutions, we have to provide appropriate value of $E_c$ as the boundary 
condition of the flow which is a conserved quantity (for our case) throughout the accretion 
regime. Thus one can write the conserved energy of the flow $E \equiv E_c \equiv E_{\rm out}$, where $E_{\rm out}$ is the specific energy of the flow at the outer accretion boundary which is simply related to the outer accretion radius $r_{\rm out}$ and the corresponding temperature at the outer radius 
(or the ambient temperature) $T_{\rm out}$. Hence, given a value of $E_c$, or equivalently for a corresponding choice of the outer boundary values (outer boundary conditions) of the inflow, i.e., values of ($r_{\rm out}, T_{\rm out}$), one would be able to compute the corresponding sonic location $r_{c}$ using Eqn. (25). Using $r_{c}$, the corresponding radial velocity and the sound speed at the sonic location would then be evaluated. These would then 
act as further boundary conditions of the flow to compute $v_r$ and $c_s$ from 
Eqns. (19) and (20). We solve the Eqns. (19) and (20) using the fourth order Runge-kutta method integrating from the sonic location $r_{c}$ in both inward and outward directions. In the next section we will analyse the transonic behaviour (i.e., to carry out the sonic-point analysis) of the accretion flow and perform the global analysis of the parameter space, for an inviscid spherically symmetric, steady adiabatic flows. 

\section{Sonic-point analysis for spherically symmetric accretion flow in the elliptical galaxy gravitational field} 

In the inviscid spherically symmetric, steady adiabatic (or polytropic) Bondi-type accretion onto the gravitational field of an isolated BH, or even for the case with isolated BH in the presence of $\Lambda$ (e.g., GB15; Mach et al. 2013), the dynamical flow behaviour of the flow is always characterized by a a single sonic transition, where the flow makes a transition from subsonic speed to supersonic speed through a single `saddle-type' sonic point (or critical point) in the region close to the central object. However, in case of the accretion flow traversing through the entire gravitational field of the elliptical galaxy onto the central SMBH, the transonic behaviour of the flow and consequently the nature of flow dynamics may substantially deviate from that of the classical Bondi-type solution. The transonic behaviour or the transonicity of the fluid flow can be determined using Eqn. (25) and/or Eqn. (26), which describe the `sonic energy-sonic radius', and `sonic entropy-sonic radius' profiles, respectively. 

In figures 1a,c,e, we show the variation of sonic energy ($E_c$) of the accretion flow as a function of sonic location ($r_c$) for our five-component elliptical galaxy (BH + stellar + DM + hot gas + $\Lambda$). Here we consider the standard value of $\Upsilon_B  =  100$, with JS-3/2 DM matter profile. Owing to the incorporation of galactic contribution to the potential through different galactic mass-components (stellar, DM, hot gas) in the presence of $\Lambda$, the accretion flow exhibits a remarkable behaviour, with the appearance of {\textit{multi-transonicity}} or {\it{multi-criticality}} in the fluid flow, as is being displayed in the $E_c - r_c$ profiles in Fig. 1. This emergence of {\it{multi-transonicity}} or {\it{multi-criticality}} in a spherically symmetric, steady adiabatic flow, indicates a significant departure from the classical Bondi solution. Although, there may be instances when more than one critical point may also appear in the spherical accretion flow onto an isolated BH, if, however, the flow deviates from being strictly adiabatic (see paragraph 5 in \S 1). To check the relative contributions of different galactic components to the transonicity of the flow, as an example, in Fig. 1a, we show the profiles corresponding to the contributions from only (BH + stellar) components as well due to contributions from (BH + stellar + DM + hot gas) components neglecting the effect of $\Lambda$. Figure 1a reveals that stellar mass component dominates the central pc to kpc-scale regions, whereas the DM and the hot gas components dominate the regions in the scales of $\sim$ few kpc to few hundreds of kpc. The effect of $\Lambda$ becomes dominant in the region $\gtrsim$ 100 kpc. 

Let us focus on Fig. 1a, corresponding to our five component case. The nature of the profile reflects that within a certain range in the value of $E_c$, i.e., from $E_{\rm min}$ to $E_{\rm max}$ (where the curve peaks), the flow comprises of three sonic locations for any particular choice in the value of $E_c$ \footnote{Note that for any particular accretion flow, $E_c$ is equivalent to the total energy $E$ of the flow.}. To exemplify, we have marked a representative line of constant energy $E_c \simeq 2.24 \times 10^{-6}$ on the figure which intersects at three sonic locations. The inner and outer sonic points with the negative slopes of the curve are the `saddle type' or `X type' sonic points through which the subsonic to supersonic transition takes place, with the matter attaining supersonic speed. On the contrary, in-between two X-type sonic points, one with the positive slope of the curve is the `centre-type' or `O-type' sonic point. $E_{\rm min}$ actually corresponds to the radial location where the profile curve gets truncated. This radial location is referred to as the static radius ($r_{\rm stat})$; the radius, at which the gravitational attraction due to galactic mass distribution is counterbalanced by the $\Lambda$ \footnote{$r_{\rm stat}$ can be determined using the limiting condition, that at $r_c > r_{\rm stat}$ as $\mathscr{F}_{\rm Gal}$ becomes negative, sonic velocities become imaginary.}. Static radius is the radius, in the vicinity of which the effect of $\Lambda$ strongly starts to dominate. Figure 1a reveals that with the increase in the value of $E_{c}$, as $E_c$ approaches $E_{\rm max}$, the outer X-type and O-type sonic points tend to coincide. And with the further increase in the value of $E_c$, for $E_c > E_{\rm max}$, the flow will only have one sonic transition through the inner X-type sonic point, resembling the classical Bondi-type solution. On the other hand, for $E_c < E_{\rm min}$, the flow comprises of only two sonic locations, with inner X-type and centre-type. Figures 1c,e are to be interpreted in a similar fashion. The figures reveal that with the increase in the value of $\Gamma$ the energy range for which one obtains three sonic locations gets reduced, with the decrease in the corresponding values of $E_{\rm max}$. It needs to be however noted that the value of $E_{\rm min}$ remains the same, as the static radius is independent of $\Gamma$. 

A similar multi-transonic behavior (as is being revealed in Fig. 1) appears in the case for advective accretion flows onto isolated BHs, which has been extensively studied in the astrophysical literature (e.g., Abramowicz \& Chakrabarti 1990; Chakrabarti 1996; Mukhopadhyay \& Ghosh 2003; Mukhopadhyay 2003). The reason for the appearance of multi-transonicity or multi-criticality in advective accretion flows is owing to the existence of centrifugal barrier in the vicinity of the BH. This centrifugal barrier occurs if the angular momentum of the flow is sufficiently high (albeit sub-Keplerian). Bondi-type spherical accretion onto the BH, on the other hand, is devoid of any such angular momentum in the flow. However, in case of the accretion flow traversing through the entire gravitational field of the elliptical galaxy onto the central SMBH, the galactic gravitational potential actually provides the necessary centrifugal barrier for the occurrence of multi-transonicity, in spherical accretion. It is to be noted that unlike the scenario in advective accretion onto an isolated BH, where the multi-transonicity appears in the region close to the BH, here, in our case, multi-transonic behaviour arises predominantly in the central to the outer regions of the accretion flow, where the galactic contribution to the gravitational potential becomes dominant.

\begin{figure*}
\centering
\includegraphics[width=150mm]{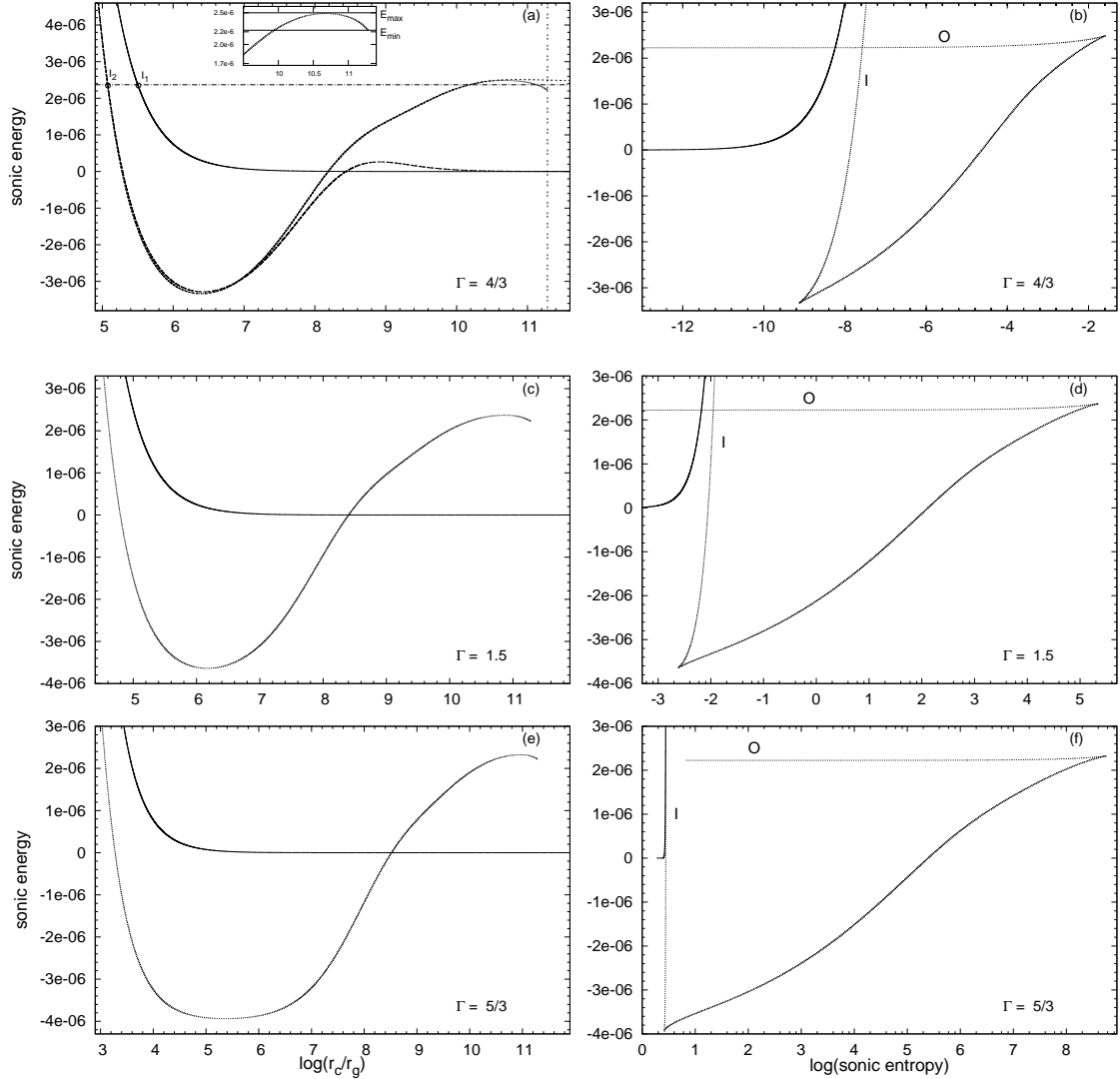}
\caption{Variation of sonic energy ($E_c$) as a function of sonic location ($r_c$) (figures 1a,c,e), and function of sonic entropy ($\dot {\mathcal{M}}_c$) (figures 1b,d,f), for spherical accretion in five-component elliptical galaxy, for various values of $\Gamma$. The curves are generated for standard value of $\Upsilon_B  =  100$, and corresponding to JS-3/2 DM profile. 
Solid curves in all the figures correspond to the profile only with the BH contribution representing the classical Bondi case. Dotted curves in all the figures correspond to the profile for our five-component case (BH + stellar + DM + hot gas + $\Lambda$). To check the relative contributions of different galactic components to the transonicity of the flow, as an example, in Fig. 1a, we show $E_c - r_c$ profiles for (BH + stellar) case (long-dashed curves) as well as for (BH + stellar + DM + hot gas) case (short-dashed curves). In figures 1b,d,f, `I' indicates inner X-type sonic point branch, whereas `O' indicates outer X-type sonic point branch, for our five-component case.   
The truncation of dotted curves in figures 1a,c,e at $r_c \simeq 1.9168 \times 10^{11} \, r_g$, indicates `static radius' $(r_{\rm stat})$. Note that $r_{\rm stat}$ is independent of $\Gamma$. As an example, in Fig. 1a, we have marked $(r_{\rm stat})$ with vertical double-dashed line. At $r_{\rm stat}$, the corresponding value of $E_c \equiv E_{\rm min} \simeq 2.2248 \times 10^{-6}$. The horizontal dotted-dashed line in Fig. 1a is a representative line of constant energy $E_c \simeq 2.24 \times 10^{-6}$ which intersects the dotted curve at three sonic locations, with two `X-type' sonic points with negative slopes, and in-between two X-type sonic points one with the positive slope of the curve is the `O-type' sonic point. $I_2$ denotes the inner X-type sonic point for five-component case, whereas $I_1$ corresponds to the sonic point for classical Bondi case. The inset image in Fig. 1a depicts the zoomed view corresponding to our five-component case in the outer regions, denoting minimum energy $E_{\rm min}$ and maximum energy $E_{\rm max}$; the energy range for which the flow will have three sonic points. {\it Rest of the panels in Fig. 1, and figures 2 and 3, are to be interpreted 
in a similar fashion}. For $\Upsilon_B  =  100$, the virial radius $r_\nu \simeq 2.554 \times 10^{10} \, r_g$. Note that for our case, $1 \, {\rm pc} \sim  10^5 r_g$. 
 } 
\label{Fig1}
\end{figure*} 

\begin{figure*}
\centering
\includegraphics[width=150mm]{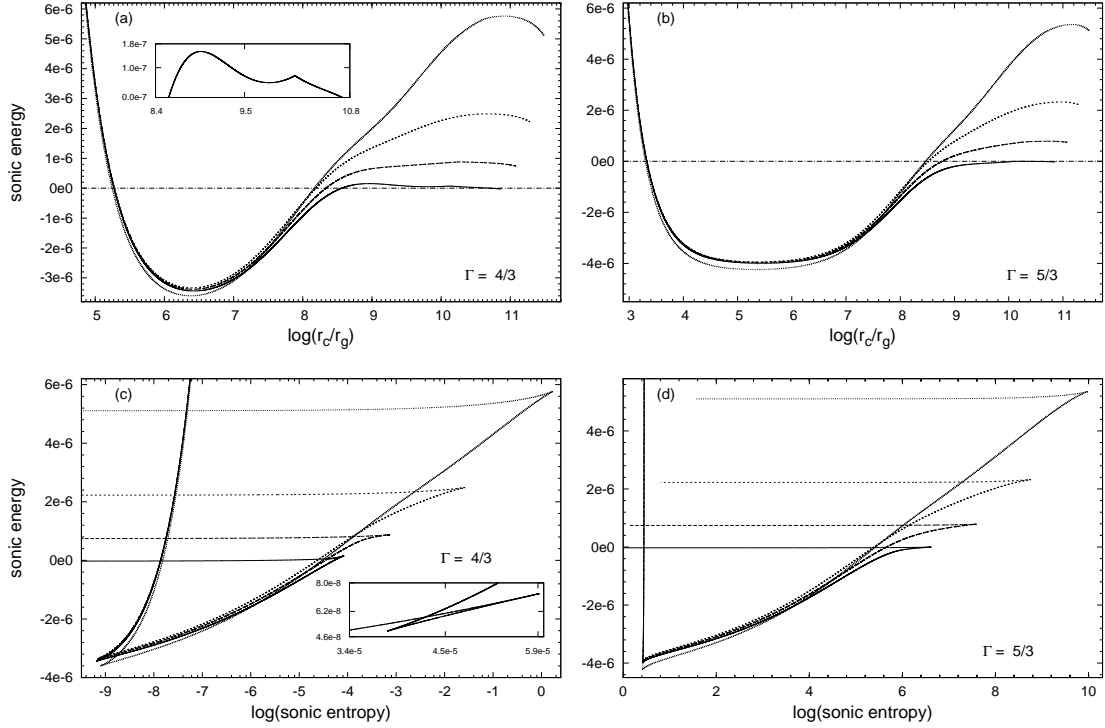}
\caption{Comparison of $E_c - r_c$ and $E_c - \dot {\mathcal{M}}_c$ profiles for different values of $\Upsilon_B$, corresponding to our five-component elliptical galaxy. 
In all the figures, dotted, short-dashed, and long-dashed curves correspond to $\Upsilon_B  \equiv (390, 100, 33)$, respectively. These curves 
are generated with JS-3/2 DM profile. Solid curves in all the figures correspond to $\Upsilon_B = 14$ for no DM case.  
The truncation of curves in figures 2a,b indicates the corresponding static radii $(r_{\rm stat})$. For the cases with $\Upsilon_B  \equiv (390, 33)$ the corresponding values of $r_{\rm stat}$ and the corresponding values of $E_c$ at $r_{\rm stat}$ [i.e., $E_{c (\rm min)}$] are \, $[r_{\rm stat}, \, E_{c (\rm min)}] \simeq [3.0792 \times 10^{11} \, r_g, \, 5.1062 \times 10^{-6}], \, \, [1.208 \times 10^{11} \, r_g, \, 7.434 \times 10^{-7}]$, respectively. The inset image in Fig. 2a depicts the zoomed view corresponding to $\Upsilon_B = 14$ in the outer regions. The radial location 
where the second peak (the more sharper) appears in this inset image is the corresponding virial radius. Similarly in Fig. 2c, the inset image depicts the zoomed view corresponding to 
$\Upsilon_B = 14$ for higher values of $\dot {\mathcal{M}}_c$. Note that for $\Upsilon_B = 14$, the corresponding $E_c$ at $r_{\rm stat}$ attains a negative value. On the other hand, for 
$\Gamma = 5/3$, corresponding to $\Upsilon_B = 14$, the $E_c - r_c$ profile shows that the corresponding values of sonic energy always remain negative for $r_c \, \gtrsim \, 2 \times 10^3 \, r_g$, implying that the flow will only have one sonic transition through the inner X-type sonic point. The virial radii for $\Upsilon_B \equiv (390, 33, 14)$ are $ \simeq (4.0202 \times 10^{10} \, r_g, \, 1.7649 \times 10^{10} \, r_g, \, 1.3262  \times 10^{10} \, r_g)$, respectively. For our case, $1 \, {\rm pc} \sim  10^5 r_g$.
 }
\label{Fig2}
\end{figure*}

\begin{figure*}
\centering
\includegraphics[width=150mm]{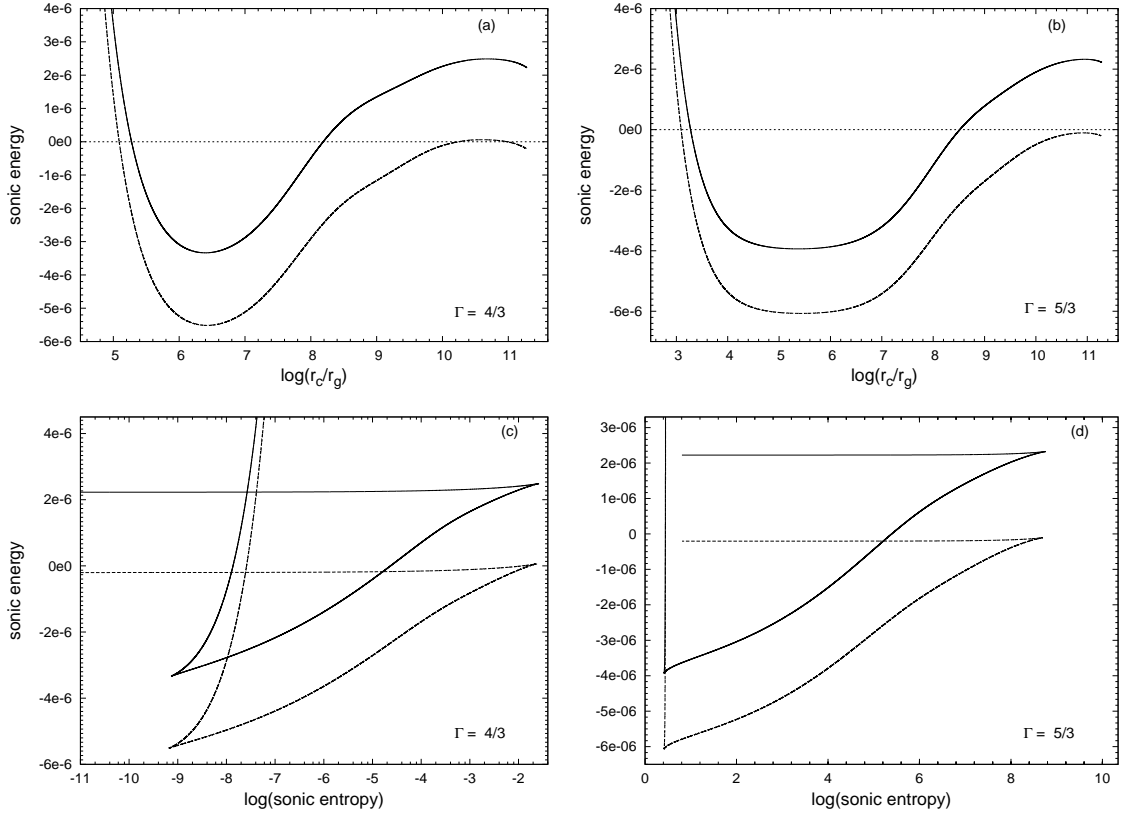}
\caption{Comparison of $E_c - r_c$ and $E_c -\dot {\mathcal{M}}_c$ profiles for JS-3/2 and NFW DM profiles, for $\Upsilon_B  =  100$, corresponding to our five-component elliptical galaxy. 
Solid and long-dashed curves in all the figures correspond to JS-3/2 and NFW DM profile, respectively. The truncation of curves in figures 3a,b indicates the corresponding 
static radii $(r_{\rm stat})$. For NFW DM case, with $\Upsilon_B = 100$, the corresponding value of $r_{\rm stat} \simeq 1.8548 \times 10^{11} \, r_g$. Note that corresponding to NFW DM profile, the corresponding $E_c$ at $r_{\rm stat}$ attains a negative value. On the other hand, corresponding 
to $\Gamma = 5/3$, for NFW DM case, the $E_c - r_c$ profile shows that the corresponding values of sonic energy always remain negative for $r_c \, \gtrsim \, 1.3 \times 10^3 \, r_g$, implying that the flow will only have one sonic transition through the inner X-type sonic point. For our case, $1 \, {\rm pc} \sim  10^5 r_g$. 
 } 
\label{Fig3}
\end{figure*} 

\begin{figure*}
\centering
\includegraphics[width=160mm]{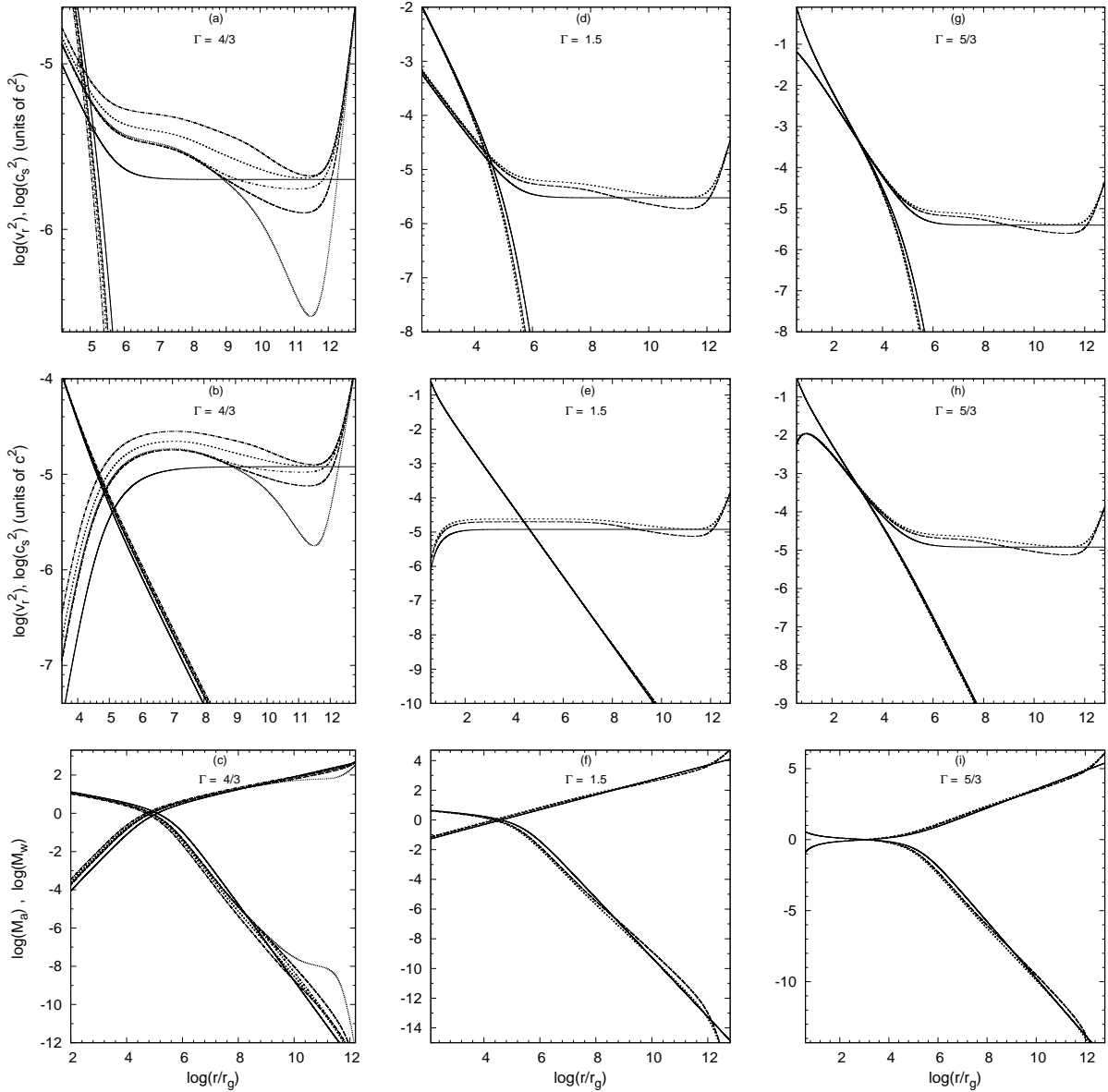}
\caption{Transonic flow profiles (radial variation) for both accretion and wind solutions with the variation of $\Gamma$ for our five-component elliptical galaxy, corresponding to various values of $\Upsilon_B$. Figures 4a,d,g correspond to accretion solution; Figures 4b,e,h correspond to wind solution; Figures 4c,f,i correspond to Mach number profiles for both accretion and wind. In Fig. 4a, thin line curves correspond to accretion sound speed, thick line curves correspond to accretion velocity: solid curve correspond to usual Bondi case; long-dashed, dotted, and short dotted-dashed curves correspond to $\Upsilon_B \equiv (100, 390, 33)$ for JS-3/2 DM case, respectively; short-dashed, and long dotted-dashed curves correspond to $\Upsilon_B \equiv (100, 390)$ for NFW DM 
case, respectively. Curves in figures 4d,g are similar to that of Fig. 4a, however for a comparison, shown only for the standard value with $\Upsilon_B  =  100$. Curves in figures 4b,e,h are similar to that of figures 4a,d,g, however, in figures 4b,e,h, the corresponding thin line curves are for wind velocity, whereas thick line curves are for sound speed corresponding to wind solution. Curves in figures 4c,f,i are similar to that of the above figures, with thin line curves are for wind, whereas thick line curves are for accretion. It needs to be mentioned that the profiles corresponding to 
$\Upsilon_B = 14$ almost coincide with that of $\Upsilon_B = 33$, we do not show them here. The transonic solutions are generated corresponding to a fixed value of $E \equiv E_c \equiv E_{\rm out} = 6 \times 10^{-6}$, as the boundary condition of the accretion flow. In Fig. 4a, the inner X-type sonic points corresponding to usual Bondi case, and corresponding to $\Upsilon_B = 390$ for NFW DM case, have been marked with circles.  
 } 
\label{Fig4}
\end{figure*}

\begin{figure*}
\centering
\includegraphics[width=160mm]{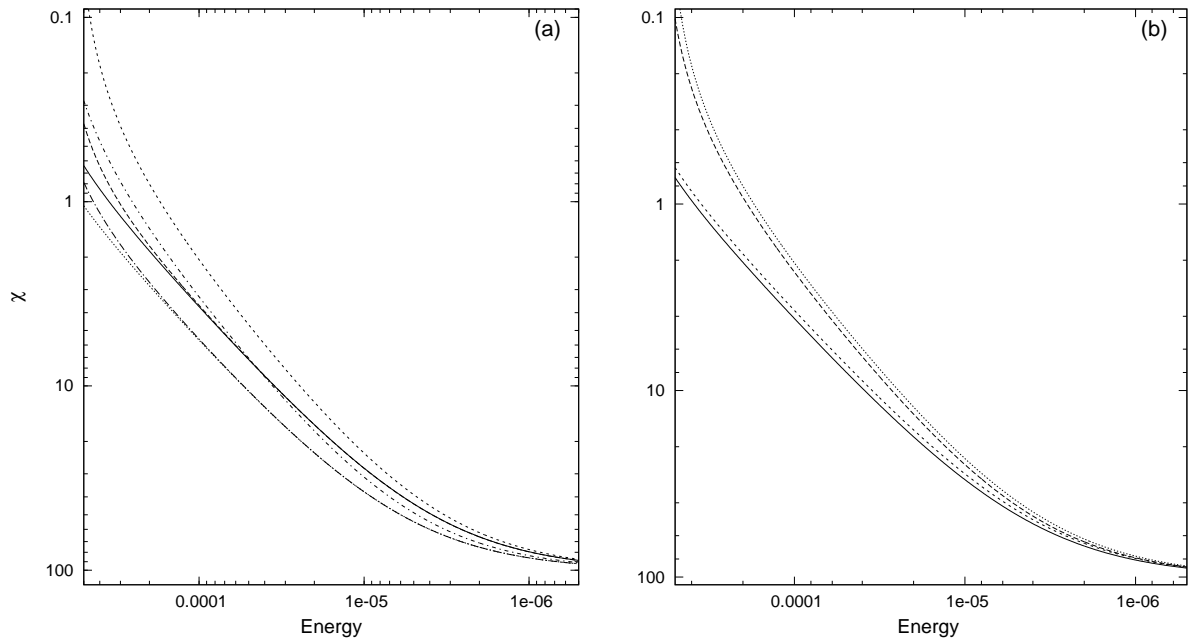}
\caption{Variation of the quantity $\chi$ with $E \equiv (E_c \equiv E_{\rm out})$, corresponding to our five-component elliptical galaxy. Solid, long-dashed, and short-dashed curves in Fig. 5a are for $\Gamma \equiv (4/3, 1.5, 5/3)$, respectively, corresponding to JS-3/2 DM profile. Similarly, dotted, long dotted-dashed, and short dotted-dashed curves are for $\Gamma \equiv (4/3, 1.5, 5/3)$, respectively, but corresponding to NFW DM profile. The profiles in Fig. 5a are generated with the standard value of $\Upsilon_B  =  100$. Figure 5b is similar to that of Fig. 5a, however generated for $\Upsilon_B \equiv (390, 33)$. For a comparison with the case for $\Upsilon_B = 100$, in Fig. 5b, the profiles are shown only for JS-3/2 DM case, for $\Gamma \equiv (4/3, 5/3)$. 
In Fig. 5b, solid and long-dashed curves are for $\Gamma \equiv (4/3, 5/3)$, respectively, corresponding to $\Upsilon_B = 390$, while short-dashed and dotted curves are for $\Gamma \equiv (4/3, 5/3)$, respectively, corresponding to $\Upsilon_B = 33$. It needs to be mentioned that the profiles corresponding to $\Upsilon_B = 14$ almost coincide with that of $\Upsilon_B = 33$, we do not show them here. 
 } 
\label{Fig5}
\end{figure*}

\begin{figure*}
\centering
\includegraphics[width=160mm]{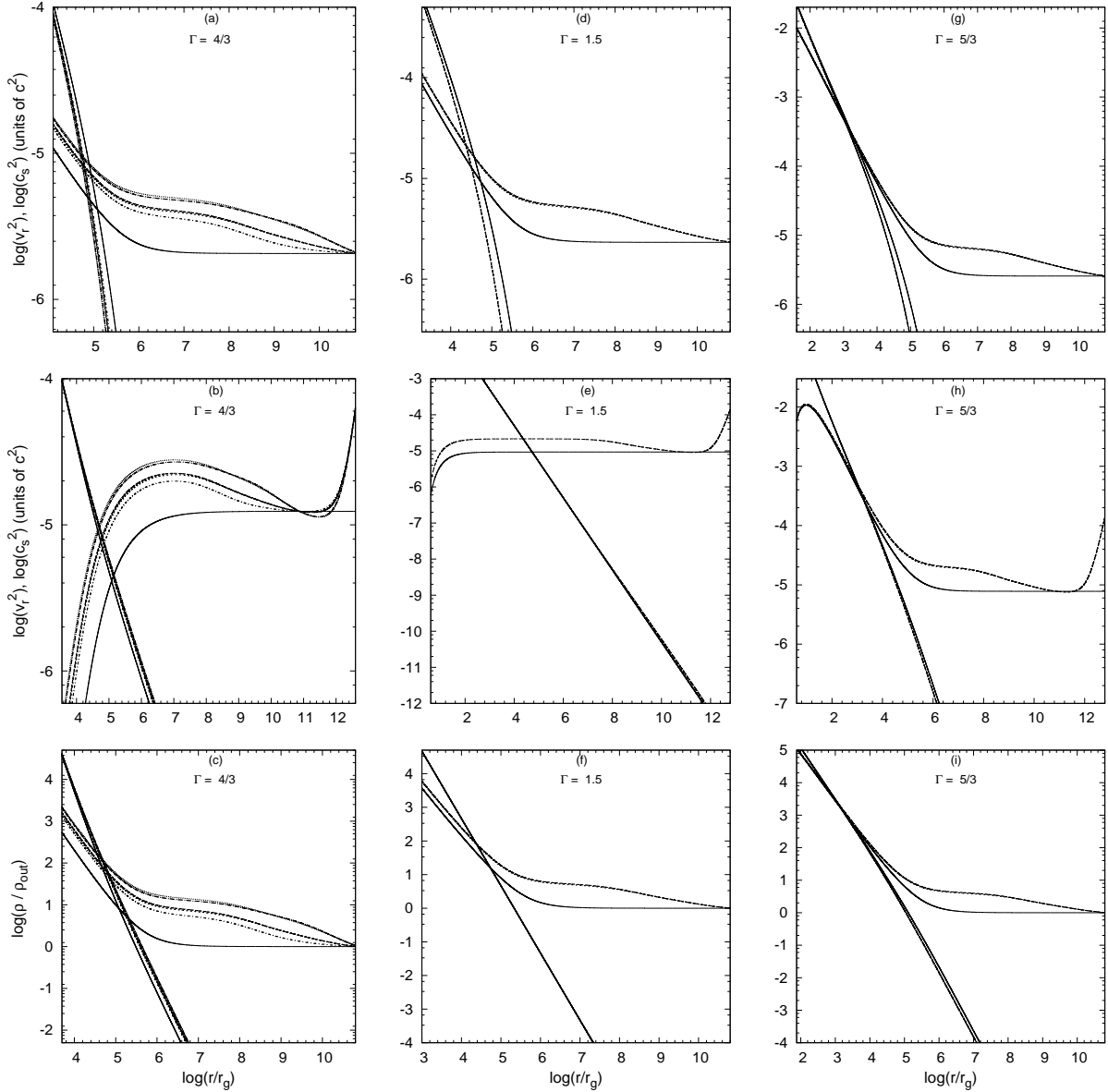}
\caption{Transonic flow profiles (radial variation) for both accretion and wind solutions with the variation of $\Gamma$ for our five-component elliptical galaxy, corresponding to various values of $\Upsilon_B$, but generated for outer accretion boundary conditions with ($r_{\rm out}, T_{\rm out}) \equiv (6 \times 10^{10} \, r_g, \, 2 \times 10^7 \, K$). Figures 6a,d,g correspond to accretion solution; (b,e,h) correspond to wind solution; (c,f,i) correspond to density ($\rho$) profiles for both accretion and wind, $\rho_{\rm out}$ represents the density at the outer boundary ($r_{\rm out}$). All the curves in figures 6a,d,g and 6b,e,h resemble the curves in figures 4a,d,g and 4b,e,h, respectively. Figures 6c,f,i are similar to that of the above figures, with thin line curves are for wind, whereas thick line curves are for accretion. It is seen that flow profiles corresponding to JS-3/2 and NFW DM cases almost coincide. 
 } 
\label{Fig6}
\end{figure*}

\begin{figure*}

\centering
\includegraphics[width=160mm]{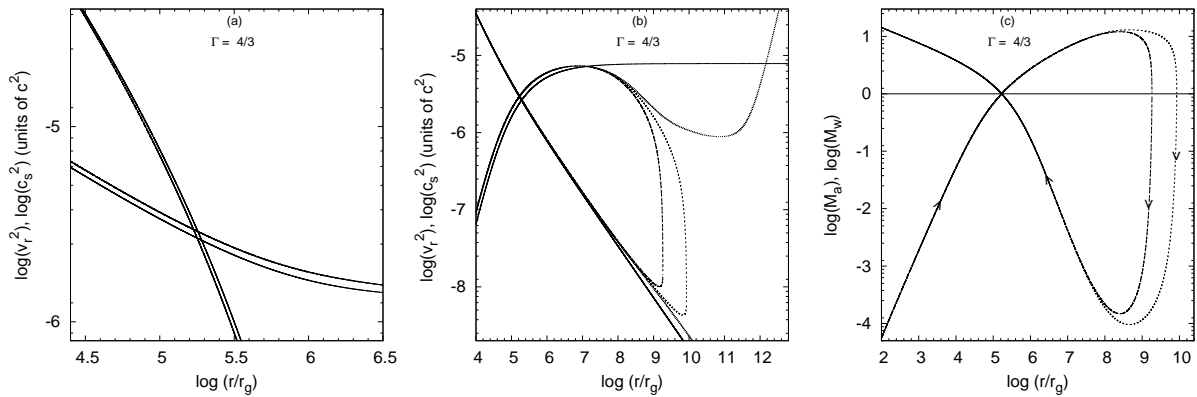}
\caption{Similar to that in Fig. 6, but generated for outer accretion boundary conditions with ($r_{\rm out}, T_{\rm out}) \equiv (\sim 10^{7} \, r_g, \, 6.5 \times 10^6 \, K$). Figures 7a,b correspond to accretion solution, and wind solution, respectively. In Fig. 7a, thick line curves correspond to accretion velocity and thin line curves correspond to accretion sound speed, whereas in 7b, thick line curves correspond to wind sound speed and thin line curves correspond to wind velocity. In both these figures, solid curve corresponds to usual Bondi flow case and long dashed and short dashed are $\Upsilon_B \equiv (100, 33)$, JS-3/2 DM case, respectively.
Dotted curve corresponds to $\Upsilon_B = 14$ for no DM case. In Fig. 7a, it is seen that the flow profiles corresponding to $\Upsilon_B \equiv (100, 33, 14)$ almost coincide. 
Here we do not show the case for $\Upsilon_B = 390$, as the flow behaviour is almost similar to that for $\Upsilon_B = 100$. 
Here the curves are generated only for $\Gamma = 4/3$, as corresponding to the stated outer accretion boundary conditions, higher values of $\Gamma = (1.5, 5/3)$ render the corresponding values of 
$E$ to be negative. Similarly is the case with NFW DM profile, for which, irrespective of the values of $\Upsilon_B$ and $\Gamma$, the stated outer accretion boundary conditions always render the corresponding values of $E$ to be negative. Figure 7c corresponds to Mach number profiles for the case with $\Upsilon_B \equiv (100, 33)$, displaying the corresponding flow topologies. Thick line curves with inward arrow direction correspond to accretion branch solution, while the thin line curves with outward arrow direction correspond to wind branch solution. The flow topologies with arrows in the curves indicate that while initially the wind flow makes an outward journey passing through the inner X-type sonic point, the matter does not find any physical path to further continue its outward journey, the matter subsequently follows the accretion branch and enters the BH. 
 } 
\label{Fig7}
\end{figure*}

\begin{figure*}
\centering
\includegraphics[width=160mm]{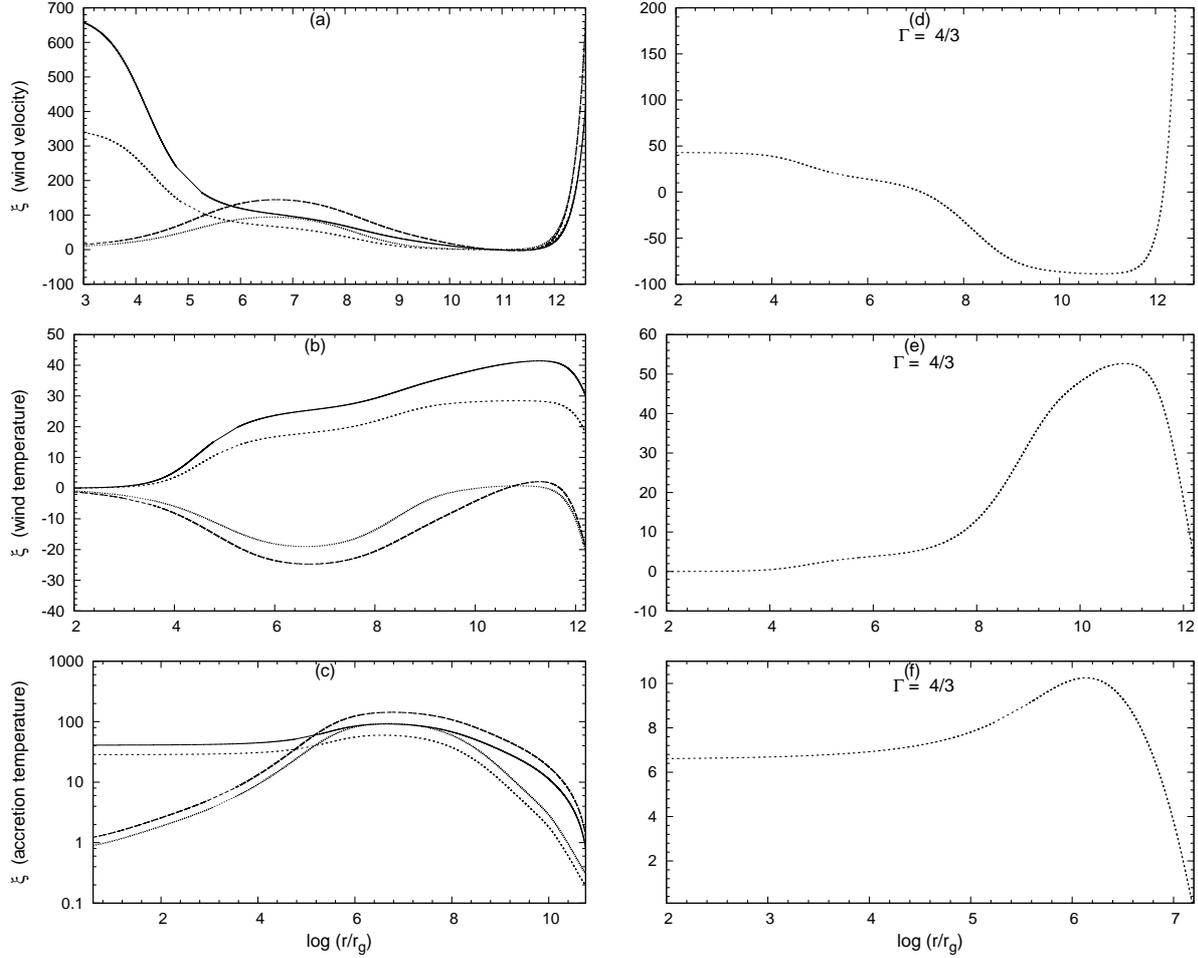}
\caption{Variation of quantity $\xi$ with radial distance $r$ for our five-component elliptical galaxy. Figures 8a,b,c are for wind velocity, wind temperature, and 
accretion temperature, respectively, corresponding to outer accretion boundary conditions with ($r_{\rm out}, T_{\rm out}) \equiv (6 \times 10^{10} \, r_g, \, 2 \times 10^7 \, K$). Figures 8d,e,f are similar to figures 8a,b,c, however corresponding to outer accretion boundary conditions with ($r_{\rm out}, T_{\rm out}) \equiv (\sim 10^{7} \, r_g, \, 6.5 \times 10^6 \, K$). For a comparison, in figures 8a,b,c, we only show the cases for standard value with $\Upsilon_B  =  100$, and for no DM case with $\Upsilon_B  =  14$, corresponding to $\Gamma \equiv (4/3, 5/3)$. In 8a,b,c solid and short-dashed curves are for $\Upsilon_B \equiv (100, 14)$, respectively, corresponding to $\Gamma = 4/3$, whereas long-dashed and dotted curves are for $\Upsilon_B \equiv (100, 14)$, respectively, corresponding to $\Gamma = 5/3$. In figures 8a,b,c, the profiles are shown for JS-3/2 DM case, as the profiles for NFW DM case almost coincide with that for the JS-3/2 DM case. In the context of figures 8d,e,f, note that corresponding to the stated outer accretion boundary conditions, for $\Gamma \equiv (1.5, 5/3)$, and NFW DM profile, one does not obtain physical solutions for accretion and wind (see Fig. 7). In figures 8d,e we show the profiles associated with the wind flow corresponding to only $\Upsilon_B  =  14$ for JS-3/2 DM case, as corresponding to other values of $\Upsilon_B$, the actual wind flow is not physically possible (see Fig. 7). Figure 8f is similar to that of figures 8d,e, but corresponding to accretion temperature (note that for accretion temperature, the corresponding profile for $\Upsilon_B = 100$ almost coincides with that for $\Upsilon_B  =  14$). The negative values in the quantity $\xi$ indicate that the corresponding values of the physical quantities for the usual Bondi flow case is larger than that for our five-component galaxy. 
 } 
\label{Fig8}
\end{figure*}

\begin{figure*}
\centering
\includegraphics[width=160mm]{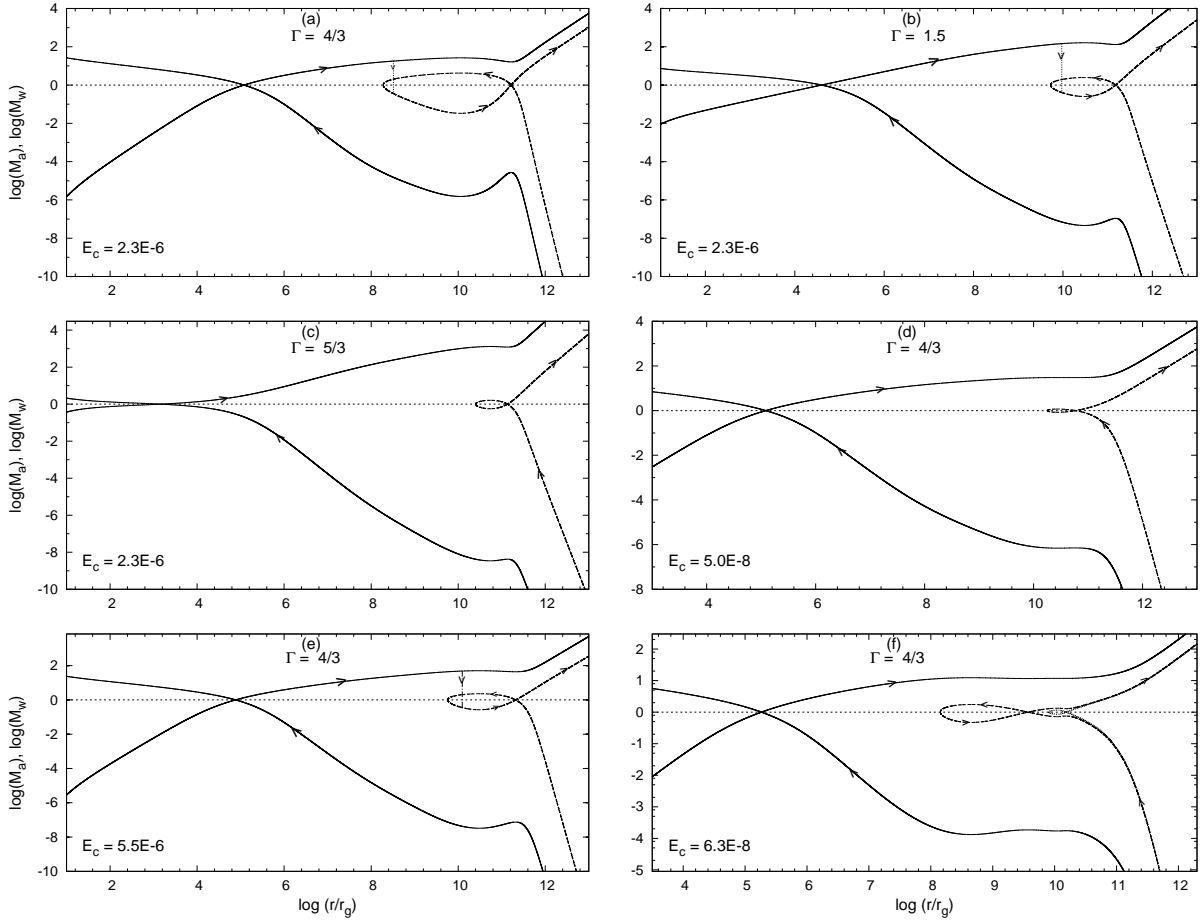}
\caption{Variation of Mach number representing global flow topologies for multi-transonic flows (with both inner and outer X-type sonic points), corresponding to the spherical accretion for our five-component elliptical galaxy, depicting a few sample cases of shock formation in the flow. Figures 9a,b,c correspond to the case with $\Upsilon_B  =  100$, for JS-3/2 DM profile. For a comparison, in Fig. 9d we show the flow topology correspond to $\Upsilon_B  =  100$, but for NFW DM profile. Similarly in Fig. 9e we show the flow topology corresponding to case with $\Upsilon_B  =  390$ for JS-3/2 DM profile. Figure 9f corresponds to $\Upsilon_B  =  14$ for no DM case, consisting of all five sonic points (three saddle-types and two centre-types). The corresponding values of $E \equiv E_c \equiv E_{\rm out}$ for which the profiles are generated are given in the respective figures. Note that for these values of $E_c$, corresponding to the respective cases in Fig. 9, the corresponding outer sonic branch (long-dashed curves) is at higher entropy than the corresponding inner sonic branch (solid curves). This can be verified from the respective $E_c - \mu_c$ profiles in figures 1,2,3. The vertical dotted lines with arrows in figures 9a,b,e indicate the respective shock transitions in the corresponding wind flows. In these respective figures, initially the wind flow passes through the inner X-type sonic point, then after encountering a shock (satisfying Rankine-Hugoniot conditions), jumps to the outer sonic branch of higher entropy along the dotted line, and then continue its outward journey passing trough the outer X-type sonic point.  
 }
\label{Fig9}
\end{figure*}

\begin{figure}
\includegraphics[width=85mm]{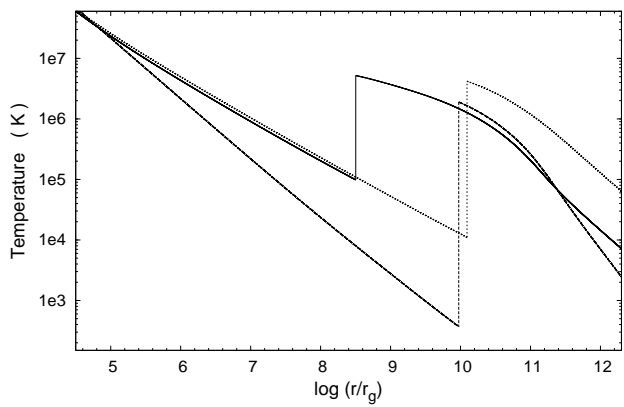}
\caption{Temperature profiles corresponding to wind flows having shocks (as described in Fig. 9). Solid, long-dashed and short-dashed curves are 
for flows having shocks corresponding to the cases in figures 9a,b,e, respectively.} 
\label{Fig10}
\end{figure}

\begin{figure*}
\centering
\includegraphics[width=150mm]{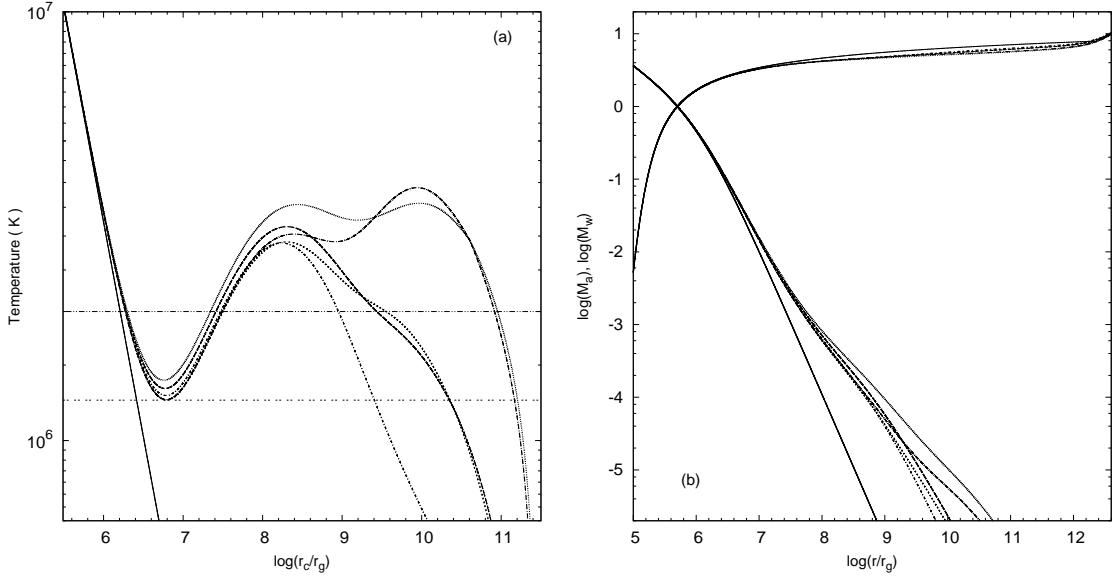}
\caption{Transonic flow behavior and flow profiles for isothermal flows, corresponding to our five-component elliptical galaxy, for different values of 
$\Upsilon_B$. Figure 11a shows the variation of temperature $T$ with sonic location $r_c$ corresponding to our five-component elliptical galaxy. Solid curve corresponds to classical Bondi flow onto an isolated BH. Long-dashed and short-dashed curves correspond to the standard value with $\Upsilon_B = 100$, for JS-3/2 and NFW DM profiles, respectively. Dotted and 
long dotted-dashed curves correspond to $\Upsilon_B = 390$, for JS-3/2 and NFW DM profiles, respectively. For a comparison, we show the profile for $\Upsilon_B = 33$, only for JS-3/2 DM case, represented by short dotted-dashed curve. The horizontal double dotted-dashed line (upper horizontal line) in Fig. 11a is a representative line of constant temperature $T \simeq 2 \times 10^{6}$ which 
intersects the corresponding curves for our five-component case at three sonic locations. The inner and outer sonic points with negative slopes of the curves are the X-type sonic points, and in-between two X-type sonic points with positive slopes of the curves are the O-type sonic points. The horizontal double-dashed line (lower horizontal line) is the representative line of constant (minimum) temperature 
$T \equiv T_{\rm min} \simeq 1.242 \times 10^{6}$ which touches the curves corresponding to $\Upsilon_B = (100, 390)$ with NFW DM profile. Below this temperature $T_{\rm min} \simeq 1.24 \times 10^{6}$, corresponding to these cases, no inner X-type sonic points appear in the corresponding $T-r_c$ profiles. In the context of this minimum temperature $T_{\rm min}$, rest of the curves for our five-component case are to be interpreted in a similar fashion. Corresponding to $\Upsilon_B = (33, 100, 390)$ with JS-3/2 DM profile, $T_{\rm min} \simeq (1.27 \times 10^{6}, 1.32 \times 10^{6}, 1.38 \times 10^{6} )$, respectively. Figure 11b depicts the Mach number profiles for both accretion and wind solutions for isothermal flows, for our five-component elliptical galaxy, corresponding to $T \sim 6.5 \times 10^6 \, K$. All the curves in Fig. 11b resemble the curves in Fig. 11a, however thin-line curves (top in the figure) correspond to wind solutions, whereas thick-line curves (below in the figure) correspond to accretion solutions. 
 } 
\label{Fig11}
\end{figure*}

\begin{figure}
\includegraphics[width=87mm]{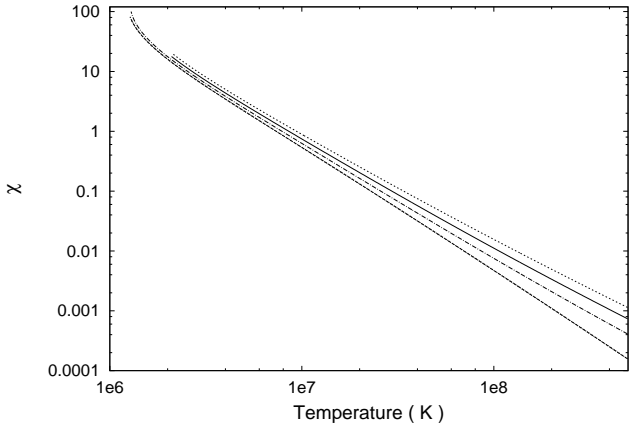}
\caption{Variation of quantity $\chi$ with $T$ for isothermal flows, corresponding to our five-component elliptical galaxy, for different values of $\Upsilon_B$. Solid and long dashed curves are for $\Upsilon_B  \equiv 100 $ for JS-3/2 and NFW profile respectively, while the short-dashed and dotted are for $\Upsilon_B  \equiv 390 $ for JS-3/2 and NFW profile respectively. For a comparison we show the profile for $\Upsilon_B = 33$, only for JS-3/2 DM case, represented by the long dotted-dashed curve. Note that the long-dashed curve and dotted curve (i,e., for NFW cases) almost coincide (the bottom-most line). 
 } 
\label{Fig12}
\end{figure}

\begin{figure*}
\centering
\includegraphics[width=160mm]{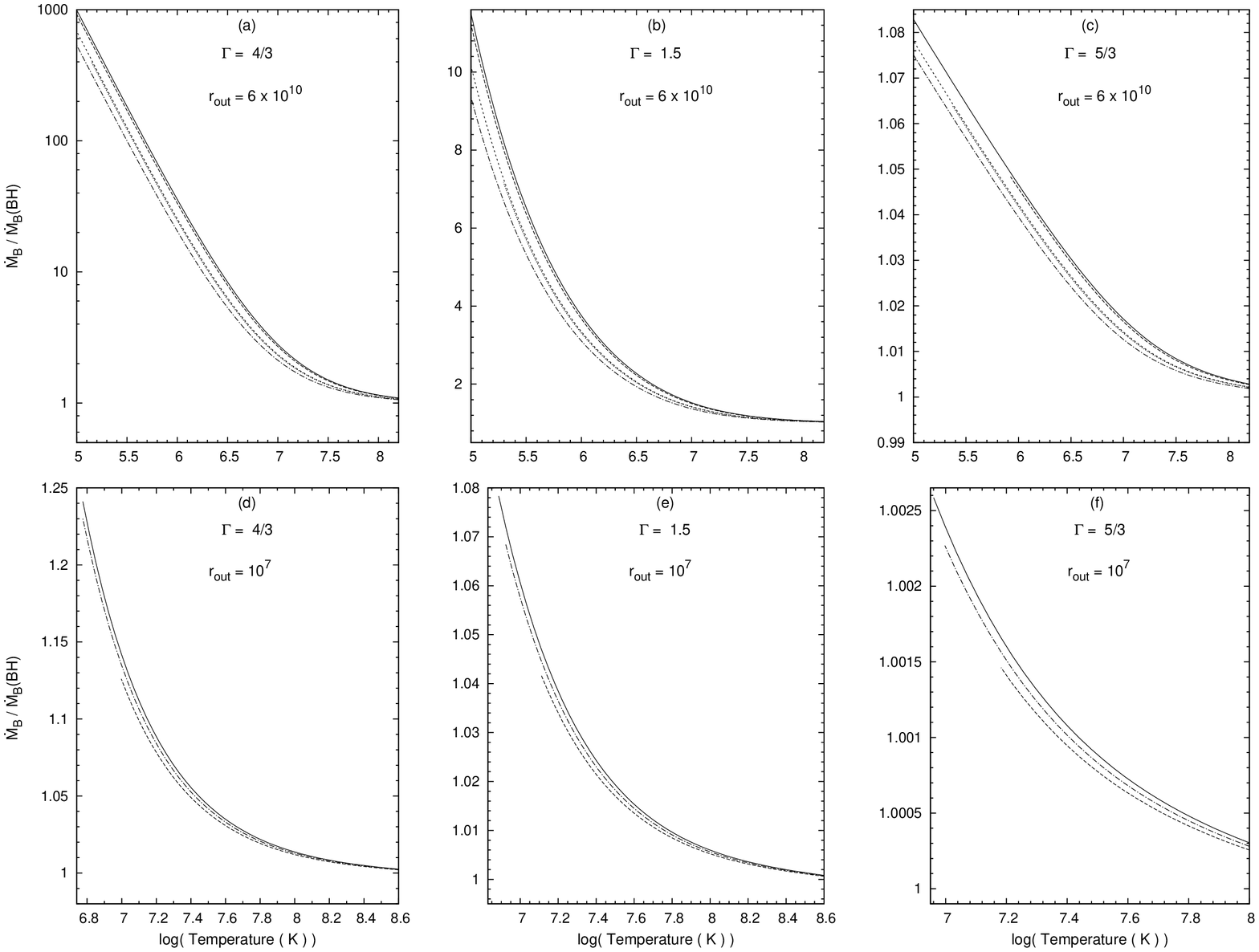}
\caption{ Variation of the quantity $\frac{\dot M_{B}}{\dot M_{B} (\rm BH)}$ with the ambient temperature $T_{\rm out}$, corresponding 
to different values of $\Gamma$. In Figures 13a,b,c the profiles are generated with the outer accretion radius $r_{\rm out} \simeq 6 \times 10^{10} \, r_g$, resembling an accretion flow from ICM/IGM, whereas figures 13d,e,f correspond to the accretion flow from the hot phase of ISM at the centre of the galaxy with 
$r_{\rm out} \simeq 10^{7} \, r_g$. In figures 13a,b,c, solid and long-dashed curves correspond to $\Upsilon_B = 100$, for JS-3/2 and NFW DM profiles, respectively; short-dashed and dotted curves 
correspond to $\Upsilon_B = 33$, for JS-3/2 and NFW DM profiles, respectively; long dotted-dashed curve correspond to $\Upsilon_B = 14$ for no DM case. We do not show the corresponding profiles for $\Upsilon_B = 390$, as their behaviour is similar to that for the case with $\Upsilon_B = 100$. For a comparison, in figures 13d,e,f, the profiles are only shown corresponding to the standard value with $\Upsilon_B = 100$, and with 
$\Upsilon_B = 14$ for no DM case, where solid and long-dashed curves correspond to $\Upsilon_B = 100$, for JS-3/2 and NFW DM profiles, respectively; long dotted-dashed 
curve in the figures correspond to $\Upsilon_B = 14$. The truncation of curves indicate that at those corresponding values of $T_{\rm out}$, the corresponding values of $E_{\rm out}$ become negative. For a comprehensive understanding of the dependence of $\dot M_{B}$ on $T_{\rm out}$, in our analysis we choose a wide range of the value of $T_{\rm out}$ in the X-axis.    
 } 
\label{Fig13}
\end{figure*}

In figures 1b,d,f we show the corresponding variation of sonic energy ($E_c$) as a function of sonic entropy ($\dot {\mathcal{M}}_c$), for our five-component elliptical 
galaxy. `I' indicates inner X-type sonic point branch, whereas `O' indicates outer X-type sonic point branch. One of the interesting implications of multi-transonicity in the fluid flow is the possible formation of shocks in the flow. Shocks can form in an accretion flow if there is a possibility of the matter to jump from the outer X-type sonic point branch of lower entropy region to the inner X-type sonic point branch of higher entropy region, whereas corresponding to wind solution, shocks can occur if there is a possibility of the matter to jump from the inner X-type sonic point branch of lower entropy region to the outer X-type sonic point branch of higher entropy region. It is seen from the figures (figures 1b,d,f) that for certain range in the value of $\Gamma$ ($\Gamma \lesssim 1.5$), inner sonic branch and the outer sonic branch intersects. The point of intersection is that point where the flow can pass through the inner and outer X-type sonic points simultaneously. This aspect of intersection of both branch points is particularly significant in the context of the formation of shocks in the flow. Nonetheless, with the increase in the value of $\Gamma$ as ($\Gamma > 1.5$), although the system shifts to higher entropy region, with the flow becoming more stable, the possibility of intersection between inner and outer sonic point branches decreases, with the outer X-type sonic points tend to shift to higher entropy region (see Fig. 1f). This indicates that although shocks can possibly occur in the flow for $\Gamma \lesssim 1.5$, however, for $\Gamma > 1.5$
the possibility of shock formation gets diminished, with the complete disappearance of possible shocks in the flow as $\Gamma \to 5/3$.

In Fig. 2 , we show the comparison of the transonic behaviour of the flow for different values of galactic mass-to-light ratio $\Upsilon_B$ corresponding to our five-component elliptical galaxy, with JS-3/2 DM profile. Figure 2 is to be interpreted in a similar fashion as that of Fig. 1. It is seen that with the increase in the value of $\Upsilon_B$, the energy range for which the 
multi-transonicity (three sonic points) occurs in the flow increases (Figures 2a,b), with the possibility of shock formation in the flow gets enhanced. It is found that 
corresponding to the case $\Upsilon_B = 14$ with negligible DM content, for a certain range of energy $E_c$, five sonic points appear in $E_c - r_c$ profile 
(see the inset image in Fig. 2a; the additional sonic points appear around the radial location where the second (more sharper) peak is visible, with the outermost sonic point, the saddle or X-type). This indicates that, in principle, there may be a possibility of having double shocks in the flow for this particular case. However, the values of energy corresponding to which multi-transonicity appears in this case are much less; the possible formation of either of the shocks in the flow is not expected to materialize. With the increase in the value $\Gamma$, as the flow tends to become nonrelativistic, the transonic behaviour of the flow corresponding to $\Upsilon_B  \equiv (390, 33, 14)$ remains largely similar as that of the case with $\Upsilon_B = 100$ (Fig. 1). In figures 2b,d, we depict the transonic behaviour of the flow for $\Gamma = 5/3$ (see the caption of Fig. 2 for more details). 

In Fig. 3, we show the comparison of the transonic behaviour of the flow for JS-3/2 and NFW DM profiles. As an example we show the comparison for standard value of $\Upsilon_B  =  100$. Figure 3 is to be interpreted in a similar fashion as that of figures 1 and 2. Although, the transonic behaviour of the flow with NFW DM profile largely resembles the case with JS-3/2 profile, however, the relevant values of energy for which one obtains three sonic points in the flow are much less as compared to that of the case with JS-3/2 profile (see the caption of Fig. 3 for more details). Note that, for 
$\Gamma = 4/3$, corresponding to NFW DM profile, although inner sonic branch and outer sonic branch intersects (Fig. 1c), however, at the point of intersection, $E_c$ remains negative. This 
indicates that corresponding to NFW DM profile, for the case with $\Upsilon_B  =  100$, the possibility of the shocks to occur in the flow is unlikely. 

Our sonic-point analysis thus reveals that owing to the galactic contribution to the potential, not only multi-transonicity appears in the flow, but also for certain parameter 
regime, depending on the values of $\Gamma$, $\Upsilon_B$ and nature of DM profile, there can be possible generation of shocks in the adiabatic spherical accretion, indicating substantial 
departure from the classical Bondi solution. Such occurrence of multi-transonicity and shocks in spherical accretion may have interesting physical 
implications, which we will discuss as we proceed through the later sections. 

In our context, when we mention shocks, we always mean Rankine-Hugoniot shocks. Rankine-Hugoniot shock conditions in 
the flow (see Chakrabarti 1989a,b; Abramowicz \& Chakrabarti 1990) are given by: 

{\scriptsize
\begin{equation}
\begin{rcases*}
\frac{1}{2} M^2_{+} \, c^2_{s+} \, + \,  N \, {c^2_{s+}} + \Psi_{\rm Gal} \, (r_{\rm sck}) \, =  \, \frac{1}{2} M^2_{-} \, c^2_{s-} \, + \,  N \, {c^2_{s-}} + \Psi_{\rm Gal} \, (r_{\rm sck}) \, , \\ 
P_{+} \, + \, \rho_{+} \, M^2_{+} \, c^2_{s+} \, = \, P_{-} \, + \, \rho_{-} \, M^2_{-} \, c^2_{s-} \, , \\
M_{+} \, c_{s+} \, \rho_{+} \, r^2_{\rm sck} \, = \, M_{-} \, c_{s-} \, \rho_{-} \, r^2_{\rm sck} \, , \\
{\dot {\mathcal{M}}}_{+}  >   {\dot {\mathcal{M}}}_{-}   \, ,       
\end{rcases*}
\label{27}
\end{equation}
}
where subscripts `$-$' and `+' symbolize the quantities just before and after the shock, respectively. $r_{\rm sck}$ represents shock location. Using first three relations of the above equation, one can compute the shock invariant quantity given by 
\begin{eqnarray} 
C = \frac{\left[1/M_{+} + \Gamma M_{+} \right]^2}{M^2_{+} \, (\Gamma -1) + 2} = \frac{\left[1/M_{-} + \Gamma M_{-} \right]^2}{M^2_{-} \, (\Gamma -1) + 2}
\label{28}
\end{eqnarray}
For the shocks to occur in the flow, it is thus necessary to simultaneously satisfy all the relations of Eqn. (27) and Eqn. (28). In the next section, we will analyse in details about the aspect of shock formation in the context of our spherical flow. 

\section{Fluid properties of spherical accretion in the elliptical galaxy gravitational field}

In the previous section we elucidated the transonic behaviour of spherical accretion in the elliptical galaxy gravitational field. Owing to the accretion flow traversing through the gravitational field of the elliptical galaxy onto the central SMBH, the galactic contribution to the potential through different galactic mass-components in the presence of $\Lambda$, significantly affect the transonic properties of the flow. In this section, we analyze the dynamical behaviour of the fluid flow in context of our spherical accretion for our five-component elliptical galaxy, to check how the galactic contribution to the potential in the presence of $\Lambda$ influence the fluid properties of the flow, in both adiabatic and isothermal paradigms, however predominantly focusing on adiabatic flows. 

\subsection{Adiabatic case}

In Fig. 4, we show the radial profiles of the flow velocity $v_r$ and the sound speed $c_s$ for both accretion and wind solutions, corresponding to different values 
$\Upsilon_B$ and $\Gamma$, for a fixed value of energy $E \equiv E_c \equiv E_{\rm out} = 6 \times 10^{-6}$. This typical choice in the value of $E$ ensures that for spherical accretion (for both accretion and wind solutions) corresponding to our five-component elliptical galaxy, the flow always makes a single sonic transition through the inner X-type sonic point, irrespective of the values of $\Upsilon_B$, $\Gamma$, and relevant DM profiles (see the figures for $E_c - r_c$ profiles in the previous section). It needs to be remembered that $E$ is simply related to the outer accretion boundary conditions ($r_{\rm out}, T_{\rm out}$). Thus corresponding to different cases considered in Fig. 4, a combination of different values of $T_{\rm out}$ and $r_{\rm out}$, would render the stated value of $E \sim 6 \times 10^{-6}$. By meticulously checking, we make a (rough) lower bound estimates of $T_{\rm out} >  7 \times 10^6 \, K$ and $r_{\rm out} >  10^{10} \, r_g$ (or equivalently $>  100 \, {\rm kpc}$); a combination of these $T_{\rm out}$ and $r_{\rm out}$ corresponding to different cases considered in Fig. 4, would approximately render a lower bound of $E \gtrsim 6 \times 10^{-6}$. This lower bound value of $r_{\rm out} \sim 100$ kpc approximately corresponds to the X-ray halo radius of giant elliptical galaxy M87 (Bahcall \& Sarazin 1977, also see Quataert \& Narayan 2000). This roughly indicates that for typical giant elliptical galaxies, this particular choice of $E$ would possibly resemble an accretion flow directly from the ICM or intergalactic medium (IGM). The profiles show that, owing to the galactic contribution to the potential, the corresponding inner X-type sonic points always shift inwards, as compared to the case for usual Bondi flow onto an isolated BH. As an example, in Fig. 4a, we have marked the inner X-type sonic points corresponding to usual Bondi case, and corresponding to $\Upsilon_B = 390$ for NFW DM case. For more clarity, one can see the $E_c - r_c$ profiles in Fig. 1. Also, with the increase in the value $\Gamma$, as the flow tends to be nonrelativistic, the corresponding inner X-type point sonic locations, 
in general, always shift inwards (as also being reflected from the figures in the previous section). The figure shows that for relativistic flows the corresponding flow profiles for different $\Upsilon_B$ as well for different DM cases, tend to spread apart, while with the increase in $\Gamma$, the corresponding profiles tend to merge with each other as well as with the usual Bondi case, indicating that the galactic contribution to the potential has more effect on relativistic flows, as compared to the flows tending to be nonrelativistic. The profiles show that the wind velocity and accretion sound speed (or equivalently the temperature of the accretion flow) substantially deviate in the outer regions (at $r > 10^{11} \, r_g  $), attaining values several orders of magnitude higher than that of the usual Bondi case, owing to the effect of $\Lambda$. On the contrary, the wind temperature substantially decreases in the outer regions as compared to that of the usual Bondi case due to the effect of 
$\Lambda$ (for clarity, in figures 4b,e,h, we do not show this region in Y-axis). Moreover, the profiles indicate that for a fixed value of energy, the galactic contribution to the potential corresponding to elliptical galaxies with higher galactic mass-to-light ratios as well as corresponding to DM profiles with steeper inner slope (e.g., JS-3/2), has greater influence on the dynamical behaviour of the flows. 

In Fig. 5, we show the percentage deviation of the inner X-type sonic locations for our five-component case from that of the usual Bondi case, denoted by the quantity $\chi$, as a function of 
energy $E$. $\chi = \frac{\vert r_{\rm in(BH)} - r_{\rm in(Gal)} \vert}{r_{\rm in(BH)}} \times 100$. Here, $r_{\rm in(Gal)}$ is the inner X-type sonic location corresponding to our five-component elliptical galaxy, whereas, $r_{\rm in(BH)}$ is the inner X-type sonic location corresponding to the usual Bondi flow onto an isolated BH. It is seen that with the increase in $E$, for all different cases considered here, the quantity $\chi$ steadily decreases. This can be understood from the fact that with the increase in $E$, the corresponding inner sonic locations, in general, always shift inwards. Owing to which, the effect of the galactic potential on the inner sonic point diminishes, while the gravitational influence of the BH on the flow increases. Note that, for a particular fixed accretion outer boundary ($r_{\rm out}$), increase in the value of $E$ implies increase in the value of outer accretion temperature or the ambient temperature ($T_{\rm out}$). The figure also indicates that NFW DM profile has a greater influence on $r_{\rm in(Gal)}$ as compared to the case with JS-3/2 DM distribution. Also it is being found that, in general, for relativistic flows, as well as for higher values of 
$\Upsilon_B$, the corresponding inner sonic locations for our five-component case shift more inwards. 

Contrary to the scenarios in Fig. 4, where the transonic flow behaviour has been studied for a fixed value of $E$, in Fig. 6, we analyze the fluid properties of the corresponding transonic solutions for 
a fixed outer accretion boundary conditions, i.e., for a fixed $r_{\rm out}$ and $T_{\rm out}$. Here, we choose typical values of the outer accretion boundary conditions as ($r_{\rm out} = 6 \times 10^{10} \, r_g, \, T_{\rm out} = 2 \times 10^{7} \, K$), resembling an accretion flow directly from the ICM/IGM. This value of $r_{\rm out} = 6 \times 10^{10} \, r_g$ \, (or equivalently $ \sim 600 \, {\rm kpc}$) may correspond to CD galaxies at the centre of rich galaxy clusters (e.g., Seigar et al. 2007; Uson et al. 1991). For e.g., IC 1101, a CD galaxy at the centre of Abell 2029 galaxy cluster, have an extended envelope whose radius may exceed $\sim 600$ kpc (e.g., Uson et al. 1991). The stated outer accretion boundary conditions would render the conserved energy of the flow $E \equiv E_c \equiv E_{\rm out}$ to be different for different values of $\Upsilon_B$, $\Gamma$, and different DM profiles, for which the flow always makes a single sonic transition through the inner X-type sonic point. It is being found that corresponding to NFW DM profile, the energy 
$E$ of the flow is always less than that corresponding to JS-3/2 DM profile, irrespective of the values of $\Upsilon_B$ and $\Gamma$. For e.g., with 
$\Upsilon_B = 100$ and $\Gamma = 4/3$, for NFW DM case $E \simeq 5.9 \times 10^{-6}$, for JS-3/2 DM case $E \simeq 8.3 \times 10^{-6}$; with $\Upsilon_B = 100$ and $\Gamma = 5/3$, 
for NFW DM case $E \simeq 3.9 \times 10^{-6}$, for JS-3/2 DM case $E \simeq 6 \times 10^{-6}$. Similarly, with $\Upsilon_B = 390$ and $\Gamma = 4/3$, for NFW DM case $E \simeq 3.62 \times 10^{-6}$, for JS-3/2 DM case $E \simeq 1.075 \times 10^{-5}$. Note that for the usual Bondi flow onto an isolated BH, for $\Gamma = 4/3$, $E \simeq 6.18 \times 10^{-6}$; for $\Gamma=5/3$, $E \simeq 3.89 \times 10^{-6}$. The profiles show that the transonic properties of the flows for the case with fixed $r_{\rm out}$ and $T_{\rm out}$ as outer accretion boundary conditions, resemble the scenarios in Fig. 4. For the wind solutions, we have extended our study beyond the accretion boundary $r_{\rm out}$, as the wind flows may well exceed the outer accretion boundary. It is being found that the $\Lambda$ has a similar effect on the wind solutions as that of the case in Fig. 4. 

In Fig. 7, we do a similar study as that corresponding to Fig. 6, however with a different outer accretion boundary conditions. Here we choose typical values of $r_{\rm out} = 10^{7} \, r_g$ 
and $T_{\rm out} = 6.5 \times 10^{6} \, K$, which would then approximately correspond to an accretion flow from the hot X-ray emitting phase of the ISM at the centre of the 
galaxy (for e.g., see NF11, where they have used similar boundary conditions). Unlike the previous boundary conditions (described in Fig. 6), for which the values of $E$ corresponding to different $\Upsilon_B$, $\Gamma$ and relevant DM profiles, always remain positive, here, however, the stated boundary conditions render the corresponding values of $E$ to remain positive only for a limited cases (see the caption in Fig. 7). With JS-3/2 DM profile, and $\Gamma=4/3$, the corresponding values of 
$E$ for $\Upsilon_B \equiv (390, 100, 33)$ are ($ 5.364 \times 10^{-7}, \, 4.878 \times 10^{-7}, \, 4.48 \times 10^{-7}$), respectively. For $\Upsilon_B = 14$, the corresponding value of $E = 4.16 \times 10^{-7}$. Note the for usual Bondi flow onto an isolated BH, for $\Gamma = 4/3, E \simeq 3.92 \times 10^{-6}$. The corresponding values of $E$ for $\Upsilon_B \equiv (390, 100, 33)$ render two sonic points (inner X-type and O-type) for the flow (also see Fig. 2), whereas for $\Upsilon_B = 14$, the corresponding $E$ generates a single inner X-type sonic location for the flow. Fig. 7 shows 
that with the stated boundary conditions, the corresponding accretion flows always find a physical path to enter the BH passing through the corresponding inner X-type sonic points. However, in the context of wind solutions, for those cases where one have two sonic points, the matter does not find any physical path to continue its outward journey. Instead, after passing through the inner X-type point the matter eventually follows the accretion branch and enters the BH (see Fig. 7c). This behaviour of spherical accretion has a very interesting consequence. While for usual Bondi flow onto an isolated BH, there would be a wide range in the values of $r_{\rm out}$ and $T_{\rm out}$, for which physical flow for both accretion and wind is possible, however, this may not be case for spherical accretion in the context of our five-component elliptical galaxy. The galactic contribution to the potential limits the range of $r_{\rm out}$ and $T_{\rm out}$, or more precisely, limits the choice of outer accretion boundary conditions for which physical flows for accretion and wind can occur. We discuss in more details on this aspect in \S 6. 

To comprehend the actual extent to which the galactic contribution to the potential in the presence of $\Lambda$ can influence the spherical accretion flow onto the central SMBH in the elliptical galaxy, we estimate the percentage deviation of various relevant physical quantities associated with spherical accretion for our five-component elliptical galaxy from that of the case for usual Bondi flow onto an isolated BH, denoted by $\xi = \frac{\mathscr{Q}_{\rm Gal} - \mathscr{Q}_{\rm BH}}{\mathscr{Q}_{\rm BH}} \times 100$. Here, $\mathscr{Q}_{\rm Gal}$ represents any arbitrary physical quantity corresponding to our five-component elliptical galaxy, whereas, $\mathscr{Q}_{\rm BH}$ represents the same physical quantity corresponding to the usual Bondi flow onto an isolated BH. In the context of spherical Bondi-type flow for our five-component elliptical galaxy, the most relevant physical quantities with which $\xi$ can be associated are wind velocity, wind 
temperature, and the corresponding accretion temperature. In Fig. 8, we show the variation of the quantity $\xi$ with the radial distance $r$ corresponding to different values of $\Upsilon_B$, $\Gamma$ and relevant DM profiles, for our five-component elliptical galaxy. We choose similar outer accretion boundary conditions as that used in figures 6 and 7, corresponding to the accretion flow from the ICM/IGM and ISM, respectively. The figure shows that the galactic contribution to the potential has a significant effect on the flow profiles. We found that apart from the substantial increase in the wind velocities in the outer regions due to the influence of $\Lambda$, the galactic potential, too, enhances the wind flow velocity in the central to inner regions of the flow. For relativistic flows, wind velocities may even get enhanced by several hundred percent in the inner regions of the flow (Fig. 8a). Also, for relativistic flows, wind temperatures may increase by $\sim 50 \, \%$ in the outer regions of the flow (Figures 8b,e). The profiles shows that in general, for relativistic flows as well as for higher values of $\Upsilon_B$, galactic contribution to the potential has greater effect on flow profiles. It is also being found that the cosmic repulsion which dominates the outer regions of the flows has an opposite effect on the velocity and the respective temperature profiles.

\subsubsection{Global flow topology and shocks} 

In the previous scenarios (described in figures 4, 6, and 7), the flow profiles have been obtained for those cases, for which we have only a single X-type sonic location (i.e, the inner X-type sonic point) for the flow. In Fig. 9, we show the flow topologies for a few sample cases, for which one can have multi-transonicity with both inner and outer saddle-types (or X-types) in the corresponding spherical flows. A very interesting scenario appears here: for certain cases, especially corresponding to moderate to higher values of $\Upsilon_B$ \, ($\Upsilon_B \, \geq \, 100$), one obtain possible Rankine-Hugoniot shocks in the corresponding wind flows, indicated by vertical dotted lines in figures 9a,b,e. However, in the context of accretion solutions, possible generation of shocks are unlikely. Nonetheless, this occurrence of shocks in the context of spherically symmetric steady adiabatic flows, is quite noteworthy, as from the point of view of a conventional scenario, it is quite unlikely to have shocks in the flows associated with adiabatic spherical accretion. In an earlier work, however, Chang \& Ostriker (1985), while investigating the spherical accretion flow with the inclusion of local heating and cooling effects with the flow deviating from being strictly adiabatic, found solutions with spherical shocks in the flow. Nonetheless, the nature of flow topology in Fig. 9 is quite similar to that of the case for advective flows onto isolated BHs, resembling their `$x\alpha$'-type trajectories (see for e.g., Abramowicz \& Chakrabarti 1990). The properties of their shock solutions show similar behaviour with that obtained corresponding to our spherical flow.  

In the advective accretion flows, the reason for the appearance of shocks in the flow close to the BH is owing to the presence of angular momentum in the system, which provides the necessary centrifugal barrier. Here, in the context of our spherical accretion, the necessary centrifugal barrier in the flow is actually provided by the galactic gravitational potential in the presence of $\Lambda$. While in case for advective flows onto isolated BHs, possible shocks appear in the region close to the BH, here, however, they possibly occur in the central to outer regions of the flows where the galactic contribution to the gravitational potential becomes dominant. Here, we do not obtain shocks in the flow for values of $\Gamma > 1.5$. This aspect of not obtaining shocks for $\Gamma > 1.5$ resembles the scenario in advective flows onto isolated BHs (e.g., Chakrabarti 1996). The non-appearance of possible shocks in advective flows for $\Gamma > 1.5$ is owing to the absence of outer X-type sonic points in the flow. In the context of our spherical flow (for our five-component case), although the outer X-type points do exist for $\Gamma > 1.5$, the possibility of intersection between outer and inner X-type points decreases (see \S 4), owing to which the possibility of shock formation gets diminished for $\Gamma > 1.5$, with the complete disappearance of possible shocks in the flow as $\Gamma \to 5/3$. 

In Fig.10, we show the temperature profiles corresponding to wind flows having shocks. The profiles show that across the shock the temperature of wind flow 
rises by orders of magnitude. This aspect of shock formation in the the central to outer regions of the wind flow may have interesting implications, possibly in the context of outflows/jets. We comment on this aspect in a greater detail in \S 7 of this paper. Nonetheless, a fact needs to 
be noted here: in the present analysis, we obtain shocks in the wind flows corresponding to the case with only JS-3/2 DM distribution profile. It would be then quite interesting to check, whether, such possibilities would also occur for other DM models available in the literature. More detailed study of global flow topology and shocks with the inclusion of other DM profiles in context of our spherical accretion will be pursued elsewhere in the near future. 

\subsection{Isothermal case} 

The equation of state of an isothermal gaseous flow is described through the relation $P = K \rho$. For an isothermal flow, the temperature or equivalently the sound speed of the accretion flow remains constant throughout the accretion regime. This then indicates that the sound speed of the accretion flow at any radii is always equivalent to the sound speed at sonic point $r_c$  (i.e., $c_{s (\rm out)} \equiv c_s \equiv c_{\rm sc}$). Hence if the temperature of the flow is known, one can compute $r_c$ using the relation \, $T \propto { r_c \, \mathscr{F}_{\rm Gal} (r_c)}/{2} $, and $r_c$ in that case will be independent of the information of the outer accretion boundary radius $r_{\rm out}$, unlike that in the case for adiabatic flows. In Fig. 11a, we show the variation of temperature $T$ of the accretion flow as a function of sonic location $r_c$, corresponding to our five-component elliptical galaxy. Here too, resembling the scenario in adiabatic flows, for certain range in the value of $T$, multi-transonicity appears in the isothermal flows. To exemplify, we have marked a representative line of constant temperature $T \simeq 2 \times 10^{6}$ which corresponds to three sonic-point scenarios. Nonetheless, unlike the scenario in adiabatic flows, where we generally obtain three sonic points, here, one may even obtain five sonic points in the flow, corresponding to higher values $\Upsilon_B$, as depicted in Fig. 11a for $\Upsilon_B = 390$. Figure 11a exhibits a notable feature: for each of the different cases corresponding to our five-component elliptical galaxy, there is a corresponding minimum value of temperature (say $T_{\rm min}$), below which, no inner X-type sonic point appears in the corresponding $T-r_c$ profiles (for details see the caption of Fig. 11a). This then indicates that for $T < T_{\rm min}$, either no physical flows would be possible, or any physical plausible flows would only pass through the outer X-type sonic points. This is in contrary to the adiabatic case, in which, the physical plausible flows towards the central object would always make a sonic transition through the inner X-type sonic point, irrespective of the nature of transonicity in the flow. In Fig. 11b, we show the Mach number profiles for both accretion and wind solutions for different values of $\Upsilon_B$ and relevant DM models, corresponding to our five-component elliptical galaxy. For our analysis, we choose the temperature of the flow $T \sim 6.5 \times 10^6 \, K$, which is a reasonable choice corresponding to inflows from ISM, as well as corresponding to inflows from ICM/IGM. The temperature of the flow is such for which one obtains a single (inner X-type) sonic point irrespective of the values of $\Upsilon_B$, through which the physical flow makes a sonic transition. 

Figure 11a reveals that contrary to the scenario for adiabatic flows, for isothermal flows, the galactic contribution to the potential renders the inner X-type sonic location to shift outwards compared to that of the usual Bondi flow case onto an isolated BH. In Fig. 12, we show the variation of the percentage deviation of the inner X-type sonic locations for our five-component elliptical galaxy from that of the usual Bondi flow case corresponding to different values of $T$ (denoted by the quantity $\chi$), as a function of flow temperature $T$, for isothermal flows. The figure shows that with the increase in the value of $T$, for all different cases considered here, the quantity $\chi$ steadily decreases, resembling a scenario for adiabatic flows. The truncation of the curves at different values of $T$ corresponding to different values of $\Upsilon_B$ and DM models, actually corresponds to those values of $T_{\rm min}$, below which no inner X-type sonic points appear in the flow. 

More detailed study of global flow topology and shocks corresponding to the isothermal case in the context of our five-component elliptical galaxy will be pursued elsewhere in the imminent future.

\section{Bondi accretion rate}

Here we investigate how the galactic contribution to the potential in the presence of $\Lambda$ influence the Bondi accretion rate. As discussed in \S 1 in the introduction, the central SMBHs in low-luminous, low-excitation radio-loud AGNs (or the jet-mode AGNs), have been widely argued to be fuelled by the hot gas from their associated hot X-ray emitting gaseous medium, through a spherical or a quasi-spherical accretion, at near-Bondi accretion rates (see the references in paragraph 3 in \S 1), with the accreting gas traversing through the host elliptical galaxy gravitational field. In these jet-mode AGNs, the total jet powers seem to correlate well with the Bondi accretion rates (e.g., Allen et al. 2006; N07). It would then be quite interesting to estimate the Bondi accretion rates corresponding to different values of $\Gamma$ for our five-component elliptical galaxy, and to compare them with that for the usual Bondi flow onto an isolated BH. In the steady state, Bondi accretion rate ($\dot M_B$) can be defined in terms of the transonic quantities of the accretion flow, given by 
\begin{eqnarray}
\vert \dot M_{B} \vert = 4 \pi r^2_c \, c_{\rm sc} \, \rho_c \, ,   
\label{29}
\end{eqnarray}
where $\rho_c$ is the density at the sonic location. Using the relation $\rho_c = \rho_{\rm out} \, \left(\frac{c_{\rm sc}}{c_{\rm out}}\right)^{2/(\Gamma -1)}$, where $\rho_{\rm out}$ represents the density of the ambient medium, i.e., the density at the outer accretion boundary $r_{\rm out}$, one can easily compute the quantity 
${\dot M_{B}}/{\dot M_{B} (\rm BH)}$, where $\dot M_{B} (\rm BH)$ denotes the usual Bondi accretion rate corresponding to the accretion flow onto an isolated BH. $\dot M_{B}$ then denotes the Bondi accretion rate corresponding to our five-component elliptical galaxy. In Fig. 13, we show the variation of the quantity ${\dot M_{B}}/{\dot M_{B} (\rm BH)}$ with the temperature of the ambient medium or the outer accretion boundary temperature $T_{\rm out}$. The profiles show that owing to the galactic 
contribution to the potential in the presence of $\Lambda$, there is an enhancement in the value of Bondi accretion rate, as compared to the corresponding 
case for an isolated BH. This is contrary to the scenario, when only an isolated BH is considered in the presence of $\Lambda$ (i.e., by ignoring the galactic mass contributions), in which case, the Bondi accretion rate gets suppressed (e.g., GB15; Mach et al. 2013). It is being found that for relativistic flows from ICM/IGM, the galactic contribution to the potential with $\Upsilon_B \, \gsim \, 100$ in the presence of $\Lambda$, may enhance the Bondi accretion rate by $\sim 10$ times corresponding to the ambient temperature $T_{\rm out} \sim 10^{6.5} \, K$, or by $\sim 50$ times corresponding to $T_{\rm out} \sim 10^{6} \, K$. A very interesting feature being found from the profiles that, corresponding to appropriate choice in the values of $r_{\rm out}$ (i.e., the location of the ambient medium), the galactic contribution to the potential limits the range in the values of $T_{\rm out}$ (or the ambient temperature), for which physical flows for accretion and wind can occur. This is being reflected from the figure where the curves get truncated. At those values of $T_{\rm out}$ where the corresponding curves get truncated, the corresponding values of $E_{\rm out}$ become negative. This precisely indicates that, corresponding to our five-component elliptical galaxy, there is a strict limit in the choice of outer accretion boundary conditions, for which, one can have Bondi-type spherical accretion onto the central BH. It is also being interestingly found that this temperature constraint is more stringent for the case when the ambient medium is located closer to the central SMBH (for instance, if the accretion flow begins from the ISM, rather than from the ICM/IGM), with the physical flow being possible only at relatively higher values of ambient temperature. The profiles also show that, in general, corresponding to NFW DM case (i.e., with less steeper inner slope) as well as with an increase in the value of $\Gamma$ (as $\Gamma \to 5/3$), the limiting values of the ambient temperature below which no Bondi-type accretion can occur, steadily increases. This aspect of the minimum ambient temperature required to trigger Bondi-type spherical accretion in the five-component elliptical galaxy can be actually attributed to the multi-transonic nature of the flow. Unlike the scenario here, for classical Bondi case onto an isolated BH, there would be wide range of ambient temperature for which the corresponding values of $E_{\rm out}$ remain positive, and consequently such minimum temperature criteria does not strictly arise in the classical Bondi case.

Let us focus on figures 13d,e,f corresponding to the spherical flows from the hot X-ray emitting phase of the ISM. Figure 13d reveals that for relativistic adiabatic flows with $\Gamma = 4/3$, corresponding to galactic mass-to-light ratio $\Upsilon_B =100$ (which is considered to be a standard value in the literature in the context of elliptical galaxy), and with JS-3/2 DM model, the value of the ambient temperature ($T_{\rm out}$) below which Bondi-type spherical accretion would not be possible is $T_{\rm out} \sim 6 \times 10^6 \, K$. From our analysis it is also being found that the corresponding profiles with universal mass-to-light ratio ($\Upsilon_B = 390$) almost coincide with that for the case with $\Upsilon_B = 100$. Again, for lower values of $\Upsilon_B$, it is found from the figures that the corresponding values of $T_{\rm out}$ below which physical flows are not possible, generally, increases. A similar observation can be made for the case with NFW DM model, however, the corresponding limiting values of $T_{\rm out}$ further increases; with $\Upsilon_B = 100$ and for $\Gamma = 4/3$, this limiting value of $T_{\rm out} \sim 10^7 \, K$. On a similar note, for nonrelativistic adiabatic flows with $\Gamma = 5/3$, corresponding to $\Upsilon_B \, \gsim \, 100$, and with JS-3/2 DM model, the corresponding limiting value of $T_{\rm out} \sim 9 \times 10^6 \, K$, whereas, with NFW DM model, this limiting value of $T_{\rm out} \sim 1.45 \times 10^7 \, K$. It needs to be noted that JS-3/2 and NFW DM profiles describe two limiting cases of a more general $(1,3,\gamma)$ DM density distribution model, described by Eqn. (9), with inner slopes of $-3/2$ and $-1$ respectively. This then indicates that, depending upon the outer boundary radius $r_{\rm out}$ of the ISM, and the value of $\Gamma$, there should be a lower-bound in the value of ambient ISM temperature, for which a Bondi-type spherical accretion onto the central SMBH can occur in an elliptical galaxy. For our case with the typical choice of $r_{\rm out} \sim 10^7 \, r_g$, we can provide a very conservative (theoretical) estimate of lower-bound limit of temperature; for relativistic adiabatic flows this limiting value of $T_{\rm out} \sim 6 \times 10^6 \, K$, and for nonrelativistic adiabatic flows the corresponding limiting value of $T_{\rm out} \sim 9 \times 10^6 \, K$, despite the fact that the hot phase of ISM at the center of galaxy may have a temperature in the range of $\sim (10^{5.5} - 10^7) \, K$. Nonetheless, several other relevant DM models do exist in the literature, and it would be quite interesting to check, whether they would provide a different (theoretical) lower-bound limit of the temperature. This is unlike the scenario when Bondi-type spherical accretion is assumed to occur only onto an 
isolated BH. Such enhanced lower-bound limits of the temperature for which a Bondi-type spherical accretion can occur, may have interesting physical implications. We comment on this in the next section.

\section{Discussion}  

LERGs, predominantly hosted by massive ellipticals, and which mostly dominate the local radio-loud population as well as centre of cool-core clusters, are being widely argued to be powered by direct hot mode accretion, with their central SMBHs fuelled directly by hot gas from their surrounding hot X-ray emitting phase through a spherical or a quasi-spherical accretion, at Bondi or at near-Bondi accretion rates. The radio-mode feedback that typically operates in these LERGs leads to mechanical heating of the surrounding gaseous medium (ISM/IGM/ICM) and prevents its radiative cooling, thereby keeping the ambient gas hot; this hot phase directly controls the fuelling of host active nucleus. The situation corresponds to a giant Bondi-type spherical/quasi-spherical accretion onto the central SMBH with the accretion flow region extending well beyond the Bondi radius exceeding several hundred pc to kpc length-scale; in the context of giant ellipticals residing at the centre of clusters, the flow region may even exceed hundreds of kpc length-scale. Thus, in the context of massive ellipticals, in addition to the gravitational field of the central SMBH, the flow is expected to be influenced by the gravitational field due to galactic mass components (stellar, DM, hot gas), along from the effect of repulsive cosmological constant $\Lambda$. 

In the context of (Bondi-type) hot mode accretion, most of the previous studies have been undertaken considering the flow onto an isolated BH (i.e., assuming a single point source of gravitation), restricting to classical Bondi solution (see \S 1 for references). In the present work, we performed a detailed study of the Bondi-type spherical accretion onto the central SMBH by incorporating 
the entire galactic contribution to the potential in the presence of $\Lambda$, considering a five-component elliptical galaxy (BH + stellar + DM + hot gas + $\Lambda$). The primary goal of this paper was to investigate, to what extent the galactic contribution to the potential impact the dynamics of spherical accretion, relative to the classical Bondi solution. Owing to the galactic contribution to the potential, the inviscid, adiabatic spherical flow displays a remarkable behavior, with the appearance of {\textit{multi-transonicity}} or {\it{multi-criticality}} in the flow, thus violating the `criticality-condition' of the classical Bondi solution. The galactic potential significantly changes the flow properties as well as the flow topology and flow structure, relative to the classical Bondi solution, with the flow topology resembling the `$x\alpha$'-type trajectories of advective accretion flows onto isolated BHs. Below we enlist some of the major findings of our study:  
\begin{enumerate}
\item Our study reveals that there is a strict lower limit of ambient temperature below which Bondi-type accretion can not be triggered. This temperature constraint is more stringent for the case when the ambient medium is located closer to the central SMBH. For instance, if the accretion flow begins from the hot phase of ISM at the centre of the galaxy, this temperature limit can be as high as $\sim 9 \times 10^6 \, K$ below which no physical plausible Bondi-type flow can occur, despite the fact that the hot phase of ISM may have a temperature in the range of $\sim (10^{5.5} - 10^7) \, K$. This implies that Bondi-type accretion may not occur from any arbitrary location (or radius) of ambient medium. It can occur from those radii, at which the ambient temperature 
should be equal to or higher than the corresponding `minimum temperature limit', indicating that the Bondi accretion strictly depends on ambient boundary conditions. This is unlike the scenario when Bondi-type accretion is assumed to occur only onto an isolated BH (i.e., for classical Bondi solution), in which case, such minimum temperature criteria does not strictly arise. 

It needs to be again remember that it is the radio-mode AGN feedback that operates in these galaxies, actually maintains the surrounding hot phase. In our context, it would then imply that the associated radio-mode feedback should maintain the required `minimum ambient temperature limit', to enable the Bondi-type accretion to occur from the corresponding ambient location. For instance, if the Bondi-type accretion needs to occur from the hot ISM-phase at the centre of the galaxy, the radio-AGN feedback should be sufficiently strong to maintain a minimum temperature of 
$\sim 9 \times 10^6 \, K$. This indicates a tight coupling between the associated radio-AGN feedback and the hot mode accretion, with the surrounding hot phase tightly regulating the fuelling of host active nucleus. Nonetheless, based on the current observations, it is hard to ascertain the precise locations of ambient medium from where Bondi-type flow can occur and the corresponding ambient 
temperatures. Unless one has a better knowledge of them, one would not be able to properly quantify the Bondi accretion rate that is required for any quantitative modelling of maintenance mode feedback.

\item Owing to the galactic contribution to the potential there is an enhancement in the value of Bondi accretion rate, relative to the classical Bondi case. This is contrary to the scenario when only an isolated BH is considered in the presence of $\Lambda$ (i.e., by ignoring the galactic mass contributions), in which case, the Bondi accretion rate gets suppressed (e.g, GB15). 
\item The galactic contribution to the potential has a significant effect on the velocity profile of spherical wind flows. We found that apart from the substantial increase in the wind velocities in the outer regions due to the influence of $\Lambda$, the galactic potential, too, enhances the wind flow velocity in the central to inner regions of the flow (Figures 8a,b). For relativistic flows, wind velocities may even get enhanced by several hundred percent in the inner regions of the flow (Fig. 8a). 
\item More notably, corresponding to moderate to higher values of $\Upsilon_B$, we obtain possible Rankine-Hugoniot shocks in the central to outer regions of the spherical wind 
flows, that is explicitly due to the effect of galactic potential, with the temperature of the wind flow across the shock region rises by orders of magnitude.

In a simplistic scenario, spherical wind flow can be visualized as spherical outflow of matter, which then corresponds to spherical expansion of gas into a vacuum. Most of the studies regarding the dynamical evolution of radio sources and their feedback energetics, is based solely on the interaction of the radio sources with the ambient gaseous medium through which they expand, without considering the explicit effect of galactic potential on their flow dynamics. As our analysis show that galactic potential significantly impact the spherical wind flow dynamics, it would be then interesting to explore, to what extent the galactic contribution to the potential in the presence of $\Lambda$ actually affects the dynamics of collimated outflows and jets, and consequently, the energetics of radio-AGN feedback. It has already been previously suggested (Stuchl\'ik et al. 2000) that $\Lambda$ can have a strong collimation effect on jets. Recently Vyas \& Chattopadhyay (2017), while investigating transonic jet flow, found internal shocks in the inner regions of jet, that are explicitly due to the effect of the geometry of the jet. The other possible shock scenario emerges, when these jets interact with the external ambient gaseous medium. A relevant question can then be put forward: can galactic potential, too, induce shocks in the jets, in similarity to that being found in spherical wind flows? If such galactic induced shocks occur in the central to outer regions of jets, it could possibly lead to abrupt deceleration of jet, thus reducing the jet terminal speed; the shocked region would be associated with strong flares, and this flaring region could be a strong site for particle acceleration. Galactic induced shocks could thus play a prominent role in controlling the dynamical evolution of radio sources, and may also be potentially linked to the observed morphological dichotomy (i.e., the Fanaroff-Riley dichotomy) in radio sources. However, this needs thorough investigation.   
\end{enumerate}

In general, our analysis shows that, for flows tending to be more relativistic, as well as for elliptical galaxies with higher galactic mass-to-light ratios, and corresponding to DM profiles with 
steeper inner slope, the galactic contribution to the potential has greater influence on the overall dynamical behaviour of the adiabatic spherical flows. For completeness, in the present study, we also briefly analysed the behaviour of isothermal spherical flows in the context of our five-component elliptical galaxy, where one again obtains multi-transonicity in the flow. However, unlike the scenario in adiabatic flows, where one generally obtains three sonic points, in the case for isothermal flows, one may even obtain five sonic points in the flow corresponding to higher values of $\Upsilon_B$. Nonetheless, more detailed study of global flow topology and shocks corresponding to isothermal case would be pursued elsewhere in the near future. Here, it needs to be pointed out that, although in the present study our choice of DM model may seem to be reasonable good that provides a good fit to simulated DM halos, however, many other DM models exist in literature which also perform relatively well in describing simulated DM halos. It would be then worthy to check whether the main findings of our work remain consistent with other relevant DM models. This could lead to explore the feasibility to 
constrain DM density profiles in giant elliptical galaxies. Such an analysis is, however, beyond the scope of the present work, and would be pursued in the near future. 

Few authors have conducted numerical simulations of magnetized spherical accretion flows. Igumenshchev \& Narayan (2002), in their three-dimensional magnetohydrodynamic simulation found that in the presence of large-scale magnetic field, the flow becomes convection-dominated that drastically changes the flow structure relative to the classical Bondi solution, and the mass accretion rate onto the BH gets significantly reduced (also see Pen et al. 2003 in this regard). On the other hand, Igumenshchev (2006), while carrying out three-dimensional simulations of spherical accretion flows with small-scale magnetic fields, concluded that stationary supersonic accretion flows cannot form in the presence of small-scale magnetic fields. Nonetheless, it would be then worthy to undertake numerical as well as analytical studies of magnetized spherical accretion in the context of our five-component galactic system, to check how the galactic potential would impact the dynamics of the magnetized spherical 
flow. Another interesting scenario would be to incorporate the information of the radiative cooling in the spherical accretion. Although such studies of spherical accretion with the inclusion of radiative processes have been performed on previous occasions (see \S 1), in recent times, Ghosh et al. (2011) found that owing to the inclusion of the effect of Comptonization, spherically symmetric Bondi flow loses its symmetry and becomes axisymmetric. It would be then quite interesting to study Bondi-type spherical accretion with the inclusion of radiative cooling in the context of our five-component elliptical galaxy, which is left for future work.  

In essence, the present study indicates that galactic potential may plausibly play an important role in controlling the dynamics of the hot mode accretion as well as outflows/jets in massive 
ellipticals. Thus, in any realistic (numerical and analytical) modelling of accretion and/or outflow/jet dynamics, one should not ignore the possible contribution of the galactic potential. This would then be expected to shed more light on the energetics of AGN feedback in the context of these massive galaxies in the contemporary Universe.

\section*{Acknowledgments}
The authors are thankful to the anonymous reviewer for providing insightful comments and suggestions to vastly improve the quality of the manuscript. 
The authors also thank Drs. Sankhasubhra Nag and Banibrata Mukhopadhyay for helpful discussions.

\end{document}